\documentclass[12pt,a4paper]{article}
\usepackage{jheppub}
\usepackage{amsmath}
\usepackage{caption}
\usepackage{subcaption}

\usepackage{url}  
\usepackage{amssymb,dsfont}  

\usepackage{flexisym}
\usepackage{breqn}
\usepackage{comment}

\usepackage{epsf,epsfig,epstopdf}
\usepackage{graphics}
\usepackage{graphicx}

\title{Two-loop $gg \to Hg$ amplitude mediated by a nearly 
massless quark
}

\preprint{{TTP16-041}}

\author[a]{Kirill Melnikov}
\author[a]{Lorenzo Tancredi}
\author[a,b]{Christopher Wever}

\affiliation[a] {
 Institute for Theoretical Particle Physics, KIT, Wolfgang-Gaede-Strasse 1, \\ 76128 Karlsruhe, Germany
}

\affiliation[b] {
 Institut f\"ur Kernphysik, KIT, Hermann-von-Helmholtz-Platz 1,\\ 76344 Eggenstein-Leopoldshafen, Germany
}

\emailAdd{kirill.melnikov@kit.edu}
\emailAdd{lorenzo.tancredi@kit.edu}
\emailAdd{christopher.wever@kit.edu}


\abstract{We analytically compute the two-loop scattering 
amplitude $gg \to Hg$ assuming that 
the mass of the quark, that mediates  the $ggH$ interaction,  is vanishingly small.
Our computation provides an important ingredient required to improve  the theoretical 
description of the top-bottom interference effect in Higgs boson production in
gluon fusion,  and to elucidate its 
impact on the Higgs boson transverse momentum distribution.  
}

\DeclareMathOperator{\Li}{Li}

\newcommand{\ep}{\epsilon}
\newcommand{\I}{\mathcal{I}}

\newcommand{\be} { \begin{equation} }
\newcommand{\ee} { \end{equation} }

\newcommand{\kk}{\kappa}
\newcommand{\un}{\rm un}
\newcommand{\uv}{\rm UV}
\newcommand{\fin}{\rm fin}

\begin{document}
\maketitle

\setlength{\unitlength}{1.3cm} 
\section{Introduction} \label{sec:intro}

The recent discovery of the Higgs boson  at the LHC  and the close 
proximity of its properties to the Higgs boson 
of the Standard Model,   strongly suggest  that electroweak 
symmetry breaking is triggered by an elementary scalar field. 
However, since the Higgs sector of the Standard Model 
is not very compelling and since there is a large number of theoretically 
appealing alternatives, experimental exploration of the  Higgs boson 
properties has  a  very high priority for elementary particle physics.  
It is expected that the results of this research program will allow 
us to sharply contrast the observed properties of the Higgs boson 
with the Standard Model expectations. 
Because the  Standard Model is a renormalizable theory,  we can  predict the expected 
 properties of the Higgs boson with a precision that is only limited  by our ability 
to perform the required computations and by the lack of understanding of non-perturbative 
phenomena that affect the outcomes of hadron collisions.  The latter issue will probably 
prevent us from doing  sub-percent precision physics for the Higgs 
couplings,  but it is irrelevant for 
studying the proximity of the Higgs couplings to their Standard Model 
values with a few percent precision. Reaching this, the few percent,  precision in theoretical predictions for Higgs 
physics  is non-trivial; it requires many ingredients including improved perturbative predictions for major  
Higgs boson production processes. Providing such predictions is the main motivation behind the 
computation reported in this paper. 

The major Higgs boson production mechanism at the LHC is 
the gluon fusion $gg \to H$.  The interaction of the Higgs boson with gluons 
is mediated by quarks; since the  Higgs Yukawa 
couplings are proportional to quark masses, the top quark provides 
the dominant contribution to the $ggH$ interaction vertex.  
Significant theoretical advances in describing this contribution 
enabled the prediction  of   the strength of the 
$ggH$ interaction in the limit of a very heavy  top 
quark $m_t \to \infty $ with a residual uncertainty 
of about $4\%$ \cite{Anastasiou:2016cez}. 
At this level of precision, many other 
effects have to be taken into account; a 
detailed discussion can be found in Ref.~\cite{Anastasiou:2016cez}.
One  such effect,  that we focus on in this paper,  is  
the  modification of the $gg \to H$ interaction strength by  loops of 
bottom quarks.\footnote{Our computation applies to any quark whose mass is 
small compared to the mass of the Higgs boson; however,  for clarity, we will refer 
to a light quark in the loop coupled to the Higgs boson as the bottom quark.}

It may sound surprising that we need to care about the bottom quark 
loop contribution. Indeed, simple power counting indicates that 
the bottom quark  contribution is suppressed relative 
to the top quark  contribution by $m_b^2/m_H^2 \sim  10^{-3}$.  However, a more careful 
analysis of the bottom quark contributions  reveals 
that it is enhanced by two powers of a logarithm $\ln (m_H^2/m_b^2) \sim 6.5$.
As the result,  the relative suppression of the  bottom quark loop relative 
to the top quark loop  becomes much weaker,   
$m_b^2/m_H^2 \ln^2 (m_H^2/m_b^2) \sim 10^{-1}$, making a detailed  
understanding of the  bottom quark 
contribution quite relevant at the few-percent precision level.  
Since the NLO QCD corrections to 
$gg \to H$ are known to be significant, it is gratifying that these 
corrections are available for an arbitrary relation between the quark mass and 
the mass of the Higgs boson \cite{Djouadi:1991tka,Spira:1995rr}.

The situation becomes more complex if we consider less inclusive quantities, 
for example the transverse momentum 
distribution of the Higgs boson or the cross section for the  
production of the Higgs boson  in association with a jet.  
In this case,  the double logarithmic enhancement  becomes 
$p_\perp$-dependent, i.e.  
some of the large $\ln (m_H^2/m_b^2)$ logarithms  turn into $\ln (p_\perp^2/m_b^2)$.
These $p_\perp$-dependent logarithms represent a serious problem 
for $p_\perp$-resummations since their origin and their 
structure in high orders of QCD are  not understood.\footnote{For a recent 
discussion of how such logarithm arise in the abelian limit and 
in the high-energy limit, see  Refs.~\cite{Melnikov:2016emg, 
Caola:2016upw}, respectively.}
In the 
absence of  a clear understanding of how to resum 
these terms, the  extent to which 
these $p_\perp$-dependent logarithms affect  the Higgs 
transverse momentum distribution was studied empirically 
\cite{Mantler:2012bj,Grazzini:2013mca,Banfi:2013eda,
Bagnaschi:2015bop}. The results of these 
studies indicated a few percent differences  in predicted 
transverse momentum  distribution of the Higgs boson, depending  
on how these non-canonical $\log (p_\perp/m_b)$ terms are treated in the 
resummed calculations. 

It is interesting to put these studies on a more solid ground. We believe 
that a  good  starting point is the   computation of the scattering amplitude 
for $gg \to Hg$ process in the approximation where the mass of the quark 
that mediates the $ggH$ interaction is treated as  the smallest 
kinematic parameter in  the problem. Indeed, such a computation will give 
us a solid perturbative result that can be used directly to improve 
the prediction for the transverse momentum distribution of the Higgs 
boson in the region $m_b \ll p_\perp$  and, 
at the same time, an interesting data point for attempting 
the resummation of the Sudakov-like logarithms described above. 
The goal of this paper is to compute the two-loop 
$gg \to Hg$ amplitude in the  approximation $m_b \to 0$. 

We  remark that the  computation of the $gg \to Hg$ 
scattering amplitude for a nearly massless 
internal quark  is  an interesting theoretical challenge. 
Indeed,  in contrast to the 
limit of a large  internal quark mass, there is no algorithmic procedure 
to expand the Feynman  diagrams that contribute 
to $gg \to Hg$ around the vanishing quark mass.  It is possible to get around 
this problem by delaying the expansion in the small quark mass  until 
it becomes clear how it can be performed. If one thinks about this problem 
keeping in mind the established technology for higher-order 
computations that  includes  {\it i}) generation of Feynman diagrams; 
{\it ii}) projection of scattering amplitude on Lorentz-invariant form factors; 
{\it iii})  reduction of contributing Feynman integrals to master integrals 
and {\it iv})  derivation and solution of the differential equations for master integrals, 
it is easy to realize that the safest point 
to start the expansion in the small quark mass  is when the
differential equations for the master integrals are about to be solved.  Indeed, since 
the differential equations contain all the information about  
the singular behaviour  of the master integrals in the limit 
of the vanishing quark mass,  it should be possible to solve these equations by expanding 
the solutions around this singular 
point without any assumptions about the behavior of the integrals in the limit $m_b \to 0$.   
Proceeding along these lines, we can  find the master integrals by  consistently neglecting all the terms
that are power-suppressed in the $m_b \to 0 $ limit  and, at the same time, keeping  all the 
non-analytic  ${\cal O}(\log m_b)$ terms. This procedure was recently 
discussed in Ref.~\cite{Mueller:2015lrx} in the context of the inclusive 
Higgs production in gluon fusion. In this paper we describe how to  generalize it to the 
case of  $gg \to Hg$.\footnote{Recently, the planar master integrals with full mass dependence
have been computed in~\cite{Bonciani:2016qxi}.}

However, we would like to stress that 
performing the expansion at the level of the differential equations  is not optimal 
since the derivation of the differential equations 
 requires the reduction to  master integrals for an {\it arbitrary}  relation 
between the quark mass and  other kinematic parameters.  As we explain in Section~\ref{sec:formfactors}, these reductions are so demanding in terms of computing resources, that their successful completion 
should not be taken for granted. 
It is clear that, for our purposes, the full reduction is an overkill since we are 
interested in the limit $m_b \to 0$; nevertheless,  
it is non-trivial to take this limit consistently at the time of the reduction.  
It will be  important  to develop 
a computational method that will allow us to do that 
and we leave this interesting problem for the future.

The paper is organized as follows. In Section~\ref{sec:notation} we 
introduce our notation,  describe the parametrization of  the scattering amplitude 
in terms of invariant form factors and explain how to apply the renormalization 
procedure to get the  finite result.  
 In Section~\ref{sec:formfactors} we describe  how the invariant form factors 
are obtained from Feynman diagrams and how the contributing integrals 
are expressed  in terms of master integrals.  The master integrals are computed 
by solving the  differential equations, as 
we discuss  in Section~\ref{sec:deqs}.
By solving the differential equations, 
we  determine the integrals up to the  integration 
constants. Some of these constants can be obtained by  imposing
regularity conditions on the solutions,  but some can not 
and have to be computed separately. We discuss a few examples  
in Section~\ref{sec:bound}.  We present results 
for the helicity amplitudes in Section~\ref{sec:ancont}, 
discuss analytic continuation in Section~\ref{sec:cont} 
and  conclude in Section~\ref{sec:conclusions}.

\section{The scattering amplitude}\label{sec:notation}
We consider the process
\be
H(p_4) \to g(p_1) + g(p_2) + g(p_3) \label{eq:decaykin}
\ee
mediated by a bottom quark loop in a theory that includes gluons, $N_f$ massless quarks 
and the bottom quark.   The masses of the bottom quark and 
the Higgs boson are denoted by  $m_b$ and $m_h$, respectively.
We define  the Mandelstam variables
\be
s = (p_1+p_2)^2\,,\quad t=(p_1+p_3)^2\,, \quad u=(p_2+p_3)^2\,,
 \ee
subject to the constraint $ s+t+u=m_h^2$,  and note that for the process Eq.(\ref{eq:decaykin}) 
all Mandelstam invariants are positive  $s>0$, $t>0$ and $u>0$.     Following Ref.~\cite{Gehrmann:2011aa}, 
we define the dimensionless variables
\be
x = \frac{s}{m_h^2}\,,\quad y = \frac{t}{m_h^2}\,,\quad z = \frac{u}{m_h^2}\,,\quad
\kk = - \frac{m_b^2}{m_h^2}.
\ee
The scattering amplitude is a  function of $x,y,z, \kappa$ with an overall multiplicative 
factor  $( -m_h^2)^{-\ep}$ per loop.  

We would like to compute the scattering amplitude in the Euclidean kinematics. This is achieved 
by taking $m_h^2 < 0$, $s<0$, $u< 0$, $t < 0$ and keeping $m_b^2 > 0$.  In terms 
of $x,y,z$, this implies that for 
\be
0<y<1,\quad 0<z<1\,,\quad  x = 1-y-z > 0,  \quad \kappa > 0, \quad m_h^2 < 0\,,
\ee
the scattering amplitude is explicitly real. 

We denote the scattering amplitude for the process~\eqref{eq:decaykin} as
\be
{\cal A}\left (p_1^{a_1},p_2^{a_2},p_3^{a_3} \right ) = f^{a_1 a_2 a_3}\;  \epsilon_1^\mu \,\epsilon_2^\nu \, \epsilon_3^\rho 
                             {\cal A}_{\mu \nu \rho}(s,t,u, m_b)\,,
\label{eq:ampl}
\ee
where $f^{a_1 a_2 a_3}$ are the $SU(3)$ structure constants and 
$\epsilon_j (a_j)$ is the polarization vector (color label)  of a gluon with momentum $p_j$, $j=1,2,3$.
Using Lorentz symmetry and gauge invariance, one can show that the scattering amplitude
${\cal A}$ is given by a linear combination of just four form factors. 
In particular, using the  transversality conditions  
$\epsilon_i \cdot p_i = 0$, $i=1,2,3$,   and imposing a cyclic gauge fixing condition
\be
\epsilon_1 \cdot p_2 = \epsilon_2 \cdot p_3 = \epsilon_3 \cdot p_1 = 0, \label{eq:gauge}
\ee
 we can write the amplitude tensor in the following way 
\begin{align}
{\cal A}_{\mu \nu \rho}(s,t,u, m_b) &= F_1\, g^{\mu \nu}\, p_2^\rho
 + F_2\, g^{\mu \rho}\, p_1^\nu 
+ F_3\, g^{\nu \rho}\, p_3^\mu
 + F_4\, p_3^\mu p_1^\nu p_2^\rho\,. \label{eq:tensampl}
\end{align}
The four form factors $F_{1,..,4}(s,t,u,m_b)$ are Lorentz scalars; they  admit a perturbative expansion
in the QCD coupling constant. The expansion of the unrenormalized form factor reads
\begin{align}
&F^{\un}_{i}(s,t,u,m_b^2) =\sqrt{\frac{\alpha_0^3}{\pi}} \; 
 \left[  F^{(1), \un}_{i}
+  \left( \frac{\alpha_0}{2 \pi}  \right) F^{(2), \un}_{i}  + \mathcal{O}(\alpha_0^2) \right],\,\;\;\; i = 1,...4,
\end{align}
where 
$ \alpha_0$ 
is the bare QCD coupling constant.

To  perform the ultraviolet (UV) renormalization of the form factors we proceed as follows. 
First,  we express the bare coupling constant and  the bare bottom quark 
mass through their renormalized 
values. Note that this also applies to the renormalization of the Yukawa coupling 
since  $g_Y = m_b /v$, where $v$ is the vacuum expectation value of the Higgs 
 field.    Second, 
each of the form factors is multiplied by the gluon field renormalization constant 
raised to an appropriate power. This is sufficient to perform the ultraviolet 
renormalization.

In practice, we 
subtract the contribution of the massless quarks to the bare coupling constant  
in the $\overline{\rm MS}$ scheme and the contribution of the $b$-quark to the bare coupling 
constant at zero momentum transfer.  This implies the following 
relation between  the bare coupling constant $\alpha_0$ and the 
renormalized one at the scale $\mu_R$, $\alpha_s = \alpha_s(\mu_R)$
\begin{equation}
\alpha_0\, \mu_0^{2 \epsilon} \; S_\epsilon = \alpha_s\, \mu_R^{2 \epsilon}\, 
\left[ 1 - \frac{1}{\epsilon} \left( \beta_0 
+ \delta_w  \right ) \; \left( \frac{\alpha_s}{2 \pi}  \right) + \mathcal{O}(\alpha_s^2)\right],
\label{eq2.10}
\end{equation}
where $S_\epsilon = (4 \pi)^\epsilon\, e^{-\epsilon \, \gamma_E}\,, \;\; \gamma_E = 0.5772..$,
$\beta_0 = 11/6\; C_A - 2/3\, T_R \, N_f$,  $C_A= N_c$ is the number of colors, $T_R = 1/2$,
$N_f$ is  the 
number of massless quark species employed in the computation and $ 
\delta_w =  -2/3\;T_R (m_b^2/\mu_R^2)^{-\epsilon}.$

The quark mass  renormalization is performed by replacing the bare quark mass with
the on-shell renormalized quark mass. Technically, this amounts to making
the following substitution in the form factors 
\begin{equation}
m_b \to m_b\left[ 1 +  \left( \frac{\alpha_s}{2 \pi}  \right)\, \delta_m + \mathcal{O}(\alpha_s^2)\; \right]\,,
\end{equation}
where 
\be
\delta_m = C_F
\left( \frac{m_b^2}{\mu_R^2}\right)^{-\epsilon} 
 \left(  -\frac{3}{2\epsilon} -2 + \mathcal{O}(\epsilon) \right),\, 
\end{equation}
and expanding them to the appropriate order in the  strong coupling constant.  We find 
$$F_j^{(1)}(m_b) \to F_j^{(1)}(m_b) +  \left( \frac{\alpha_s}{2 \pi}  \right)\, m_b\,\delta_m \;
 \frac{d\, F_j^{(1)}(m_b)}{dm_b} 
+ \mathcal{O}(\alpha_s^2)\,.$$
Finally, the gluon wave 
function renormalization is performed by multiplying every form factor by
$$Z_A^{3/2} = 
\left( 1 + \left( \frac{\alpha_s}{2 \pi}  \right)\,
\frac{\delta_w}{\epsilon}  + \mathcal{O}(\alpha_s^2)\right)^{3/2}
= 1 + \frac{3}{2} 
\left( \frac{\alpha_s}{2 \pi}  \right)\, \frac{\delta_w}{\epsilon} + \mathcal{O}(\alpha_s^2)\,, $$
with $\delta_w$ defined after Eq.(\ref{eq2.10}), and expanding in $\alpha_s$.
With these notations, the UV-renormalized form factors become
\begin{equation}
F_j^{\uv}(s,t,u,m_b) = \sqrt{\frac{ \alpha_s^3}{\pi\, S_\epsilon^3}}\,  
\left[  F_j^{(1),\uv}
+  \left( \frac{\alpha_s}{2 \pi}  \right) F_j^{(2),\uv}  + \mathcal{O}(\alpha_s^3) \right]\,,
\end{equation}
with 
\begin{align}
 F_j^{(1),\uv} &=  F_j^{(1),\un}\,, \nonumber \\
 F_j^{(2),\uv} &= S_\epsilon^{-1}\, F_j^{(2),\un}  
 - \frac{3 \,\beta_0}{2\,\epsilon} \; F_j^{(1),\un} 
 + \, m_b\; \frac{dF_j^{(1),\un}}{d m_b}\, \delta_m  \label{eq:UVren}
 \,.
\end{align}

The UV-renormalized form factors still contain poles in $\epsilon$ that are of soft and collinear  origin.
For a generic NNLO QCD scattering amplitude, 
it was shown in Ref.~\cite{Catani:1998bh} that all such 
poles can be written in terms of tree- and one-loop amplitudes of a given process. 
Although we perform a two-loop computation, the one-loop amplitude for $gg \to Hg$ is the leading term 
in the perturbative expansion; for this reason, it is sufficient to use the NLO QCD
results of 
Ref.~\cite{Catani:1998bh},   
to isolate the infra-red and collinear poles. We therefore write
\be
F_j^{(1), \fin} = F_j^{(1), \uv}\,,\;\;\;\;\;
F_j^{(2), \fin} = F_j^{(2),\uv} - I_1(\epsilon)  F_j^{(1),\uv}\,, \label{eq:IRsub}
\ee
where in the case of three external gluons the $I_1(\epsilon)$ operator reads
\begin{equation}
I_1(\epsilon) = -\frac{ C_A e^{\epsilon \gamma}}{2 \Gamma(1-\epsilon)}
\left( \frac{1}{\epsilon^2} + \frac{\beta_0}{C_A} \frac{1}{\epsilon}\right)
\left( \left(-\frac{s}{\mu_R^2}\right)^{-\epsilon} + \left(-\frac{t}{\mu_R^2}\right)^{-\epsilon}
 + \left(-\frac{u}{\mu_R^2}\right)^{-\epsilon} \right)\,.
\label{eq:cataniI1}
\end{equation}

We  
perform the UV renormalization at the 
scale $\mu_R^2 = m_h^2$. 
We verified explicitly that the IR poles in the form factors are 
removed by the  subtraction in Eq.~\eqref{eq:IRsub}.

\section{Calculation of the form factors}
\label{sec:formfactors}

The direct computation of the decay amplitude ${\cal A}$, using the standard methods for  
multi-loop computations, 
is difficult because the amplitude depends on the polarization vectors of the external gluons. 
We can get around this problem  by computing the form factors  instead. 
To this end,  we design projection operators to extract  contributions of 
different Feynman diagrams to the four form factors. 
We define four projection operators  $P_j^{\mu \nu \rho}$ by requiring that they satisfy 
the following equation
\begin{align}
\sum_{pol}\, P_j^{\mu \nu \rho}\, 
(\epsilon_1^{\mu})^* \epsilon_1^{\mu_1}\;
(\epsilon_2^{\nu})^* \epsilon_2^{\nu_1}\;
(\epsilon_3^{\rho})^* \epsilon_3^{\rho_1}\; {\cal A}_{\mu_1 \nu_1 \rho_1}(s,t,u, m_b)  = F_j(s,t,u,m_b)\,,
\label{eq:projdef}
\end{align}
where, for consistency with Eq.\eqref{eq:gauge},  sums over polarizations of  
external gluons are  taken to be 
\begin{align}
&\sum_{pol} \left( \epsilon_1^{\mu}(p_1) \right)^* \epsilon_1^{\nu}(p_1) = 
- g^{\mu \nu} + \frac{p_1^\mu p_2^\nu + p_1^\nu p_2^\mu}{p_1 \cdot p_2}\,, \label{eq:polsums1}\\
&\sum_{pol} \left( \epsilon_2^{\mu}(p_2) \right)^* \epsilon_2^{\nu}(p_2) = 
- g^{\mu \nu} + \frac{p_2^\mu p_3^\nu + p_2^\nu p_3^\mu}{p_2 \cdot p_3}\,,  \label{eq:polsums2}\\
&\sum_{pol} \left( \epsilon_3^{\mu}(p_3) \right)^* \epsilon_3^{\nu}(p_3) = 
- g^{\mu \nu} + \frac{p_1^\mu p_3^\nu + p_1^\nu p_3^\mu}{p_1 \cdot p_3} \,. \label{eq:polsums3}
\end{align}
We stress at this point  that all Lorenz indices in Eq.(\ref{eq:projdef}) 
have to be understood  as  $d$-dimensional. 
The explicit form of the projection operators can be found by  making  an Ansatz in terms
of the same linearly independent tensors as in Eq.\eqref{eq:tensampl}
\begin{align}
P_j^{\mu \nu \rho} &= \frac{1}{d-3} 
\left [ c_1^{(j)}\, g^{\mu \nu}\, p_2^\rho
 + c_2^{(j)}\, g^{\mu \rho}\, p_1^\nu + c_3^{(j)}\, g^{\nu \rho}\, p_3^\mu
 + c_4^{(j)}\, p_3^\mu p_1^\nu p_2^\rho\, \right], 
\end{align}
where $ j \in \{1,2,3,4 \}$.  The scalar functions 
$c_i^{(j)}$ are unknown a priori; they are found 
by requiring  that Eq.\eqref{eq:projdef} is satisfied.  We obtain 
\be
\begin{split}
&c_1^{(1)} = \frac{t}{s\,u}\,, \qquad
c_2^{(1)} = 0 \,, \qquad 
c_3^{(1)} = 0\,, \qquad 
c_4^{(1)} = -  \frac{1}{s\,u}\,, 
\\
&c_1^{(2)} = 0\,, \qquad
c_2^{(2)}=  \frac{u}{s\,t}  \,,\qquad
c_3^{(2)} = 0\,, \qquad 
c_4^{(2)} = -  \frac{1}{s\,t}\,,
\\
&c_1^{(3)}= 0\,, \qquad
c_2^{(3)}= 0 \,, \qquad
c_3^{(3)} =  \frac{s}{t\,u}\,, \qquad
c_4^{(3)}= -  \frac{1}{t\,u}\,,
\\
&c_1^{(4)} =  -\frac{1}{s\,u}\,,  \quad
c_2^{(4)} =  - \frac{1}{s\,t}\,, \quad 
c_3^{(4)} =  - \frac{1}{t\,u}\,, \quad
c_4^{(4)} =  \frac{d}{s\,t\,u}\,.
\end{split} 
\ee

\begin{figure}[t!]
\centering
\includegraphics[width=0.35 \linewidth]{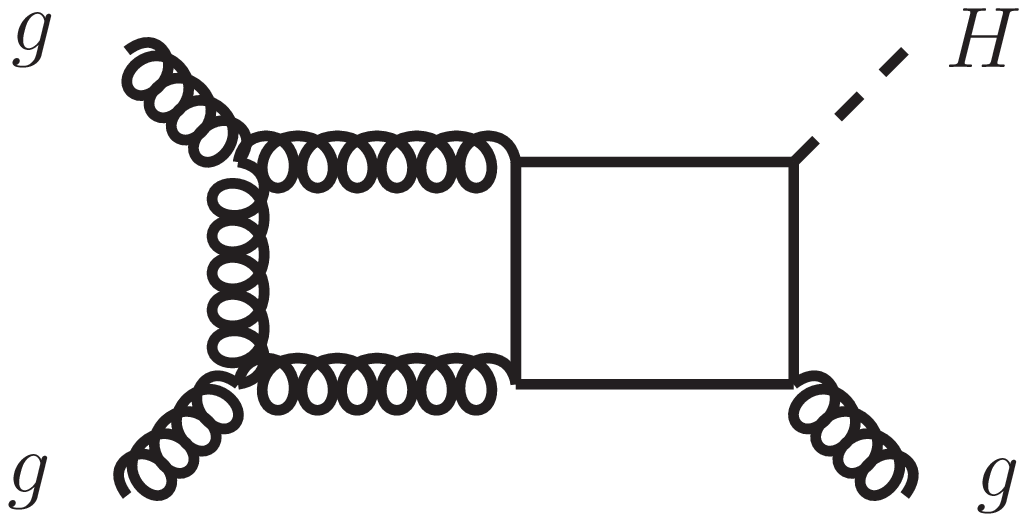} \hspace{0.4 cm}
\includegraphics[width=0.35 \linewidth]{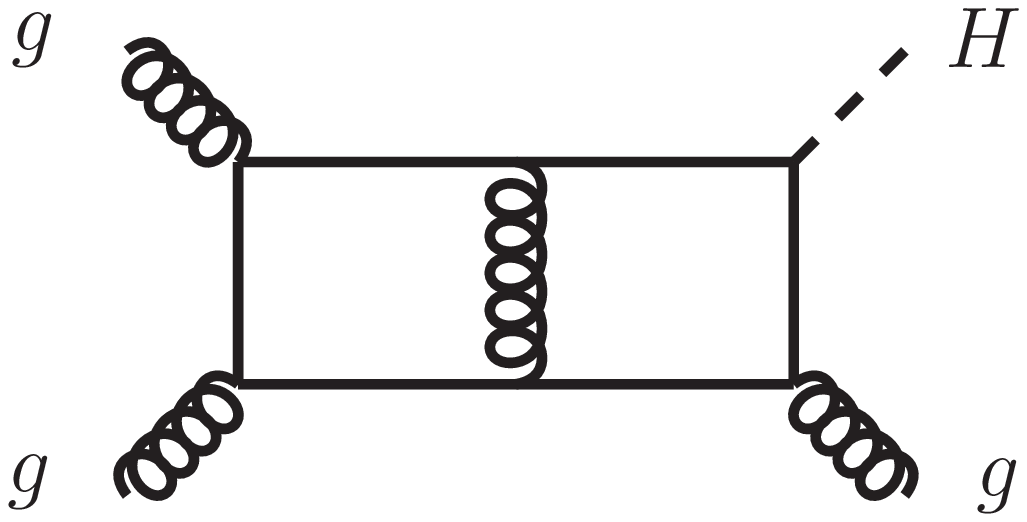} \\
\includegraphics[width=0.35 \linewidth]{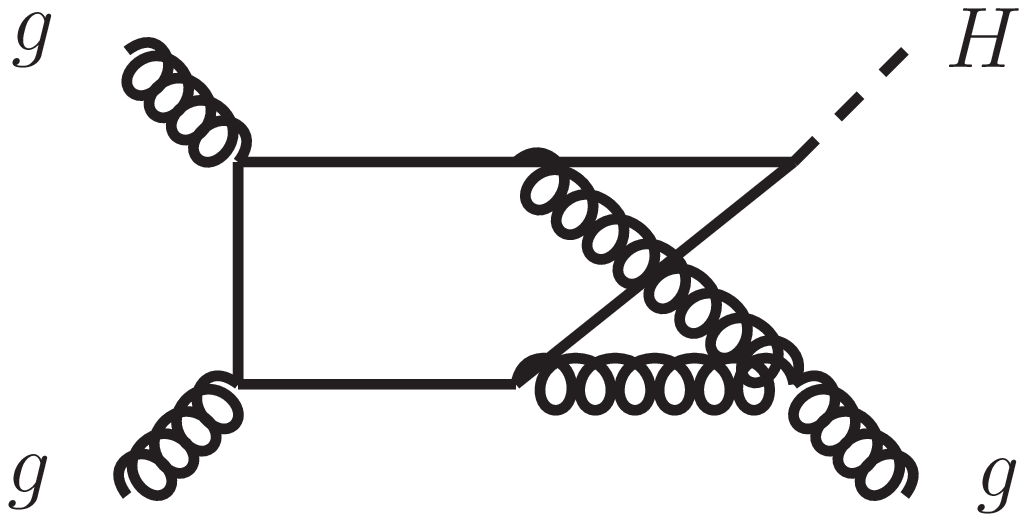} \hspace{0.4 cm}
\includegraphics[width=0.35 \linewidth]{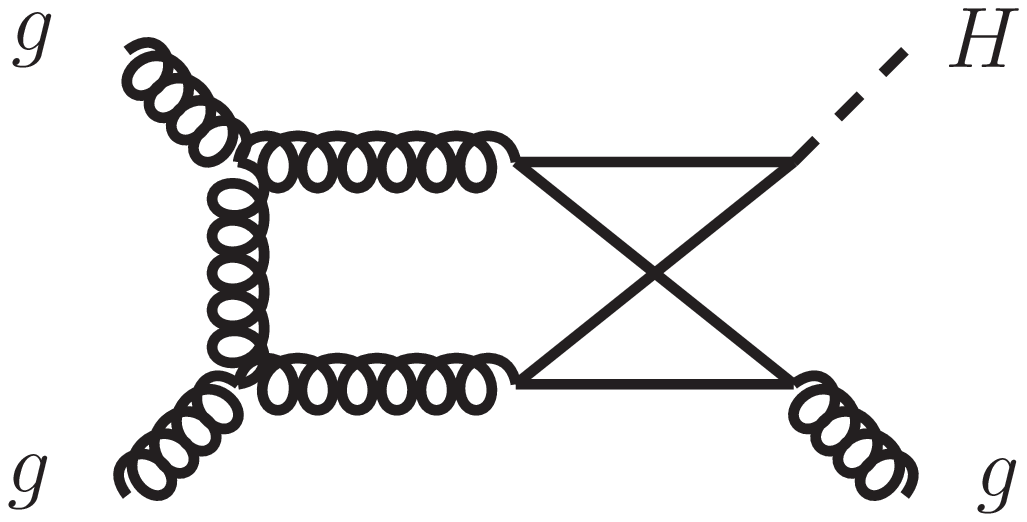} 
\caption{Examples of two-loop Feynman diagrams that contribute to the process $gg\rightarrow Hg$.}
\label{fig::feyndiag}
\end{figure}

With these results at hand, we can compute each of the form factors separately. 
Since the form factors are independent of the external polarization vectors, 
all the standard techniques employed for 
 multi-loop computations can be applied. 
In practice, we proceed as follows.  
We generate the relevant one- and two-loop Feynman diagrams with 
QGRAF~\cite{Nogueira:1991ex}. A few examples of the two-loop Feynman 
diagrams that contribute to the $gg \to Hg$ amplitude are shown in Fig.~\ref{fig::feyndiag}. 
The projection operators are applied diagram by diagram and
the polarization sums  are computed following 
Eqs.(\ref{eq:polsums1},~\ref{eq:polsums2},~\ref{eq:polsums3}).
 Once this step is completed,  each contributing  diagram
is written in terms of integrals that depend on the scalar products of the loop momenta between 
themselves  and  the scalar products of the loop momenta with  
the external momenta.  We can assign  all 
Feynman integrals that contribute to the scattering amplitude to  three integral 
families, two planar and one non-planar. These integral families are given by 
\be
\I_{\rm top}(a_1,a_2,...,a_8,a_9)
= \int 
\frac{\mathfrak{D}^dk \mathfrak{D}^dl}{[1]^{a_1} [2]^{a_2}  [3]^{a_3}  [4]^{a_4}  [5]^{a_5}  [6]^{a_6}  [7]^{a_7}  [8]^{a_8}  [9]^{a_9} 
},
\label{eq3.5a}
\ee
where ${\rm top} \in \{{\rm PL1, PL2, NPL}\}$ is the topology label and the propagators 
$[1],[2],...,[9]$ for each topology are shown in Table~\ref{tab:topos}. The
integration measure is defined as 
\be
\mathfrak{D}^dk = (-m_h^2)^{(4-d)/2}\frac{(4\pi)^{d/2}}{i\Gamma(1+\epsilon)} \int \frac{d^dk}{(2\pi)^d}.
\ee

\begin{table}[t]
\begin{center}
\begin{tabular}{| c | l | l | l|}
\hline
Prop. & Topology PL1 & Topology PL2 & Topology NPL \\
\hline
$[1]$ & $ k^2$ &                                  $k^2-m_b^2$  &                         $k^2-m_b^2$ \\
$[2]$ & $(k-p_1)^2$ &                         $(k-p_1)^2-m_b^2$  &               $(k+p_1)^2-m_b^2$ \\
$[3]$ & $(k-p_1-p_2)^2$ &                   $(k-p_1-p_2)^2-m_b^2$  &        $(k-p_2-p_3)^2-m_b^2$ \\
$[4]$ & $(k-p_1-p_2-p_3)^2$ &           $(k-p_1-p_2-p_3)^2-m_b^2$  & $l^2-m_b^2$ \\
$[5]$ & $l^2-m_b^2$ &                         $l^2-m_b^2$  &                         $(l+p_1)^2-m_b^2$ \\
$[6]$ & $(l-p_1)^2-m_b^2$ &                $(l-p_1)^2-m_b^2$  &                $(l-p_3)^2-m_b^2$ \\
$[7]$ & $(l-p_1-p_2)^2-m_b^2$ &         $(l-p_1-p_2)^2-m_b^2$  &        $(k-l)^2$ \\
$[8]$ & $(l-p_1-p_2-p_3)^2-m_b^2$ &  $(l-p_1-p_2-p_3)^2-m_b^2$  & $(k-l-p_2)^2$ \\
$[9]$ & $(k-l)^2-m_b^2$ &                     $(k-l)^2$  &                               $(k-l-p_2-p_3)^2$ \\
\hline 
\end{tabular}
\caption{Feynman propagators of the three integral families. } \label{tab:topos}
\end{center}
\end{table}

We note that the loop momenta shifts required to map contributing Feynman diagrams on to
the integral families  are obtained using the shift finder implemented in 
Reduze2~\cite{vonManteuffel:2012np}.
All algebraic manipulations needed at different stages of the computation are performed using
FORM~\cite{Vermaseren:2000nd}.
Once the amplitude is written in terms of scalar integrals, we simplify them using
all possible loop momenta shifts with a unit Jacobian; this 
can also be done using the momentum 
shift finder of Reduze2. When the contributions of all diagrams to the form factors are summed up, 
significant simplifications occur; for example, 
only integrals with up to three scalar products are left, although 
some individual diagrams receive contributions from integrals with 
up to four
scalar products. 

Having determined all scalar integrals that contribute to the amplitude, we need to reduce them 
to master integrals.  The reduction procedure relies on a systematic application of the so-called  
integration by parts  identities (IBPs)~\cite{Tkachov:1981wb,Chetyrkin:1981qh} to the integrals 
that belong to the
three topologies defined  in Table~\ref{tab:topos}. 
This procedure is automated so that, as a matter of principle, 
  one can use the publicly available programs Reduze2~\cite{Bauer:2000cp,Studerus:2009ye,vonManteuffel:2012np,fermat}, 
FIRE5~\cite{Smirnov:2008iw,Smirnov:2014hma} and LiteRed~\cite{Lee:2012cn} to perform the reduction.  However, 
in practice,  the reduction to master integrals appears to be very challenging, 
due to the presence of a mass parameter in some of the propagators.  We stress that, 
although we will eventually obtain the result for the amplitude assuming 
that the mass parameter is small, we retain 
the full mass dependence at the intermediate stages of the computation, 
including the reduction to master integrals. 

We  have found that the publicly available reduction programs and, in some cases, also their 
private versions,  were unable to successfully complete the reduction of the
most complicated non-planar 7-propagator integrals. In order to reduce those integrals, 
we wrote   a  FORM program that  produces  and solves the IBPs for the 7-propagator non-planar 
integrals, thereby reducing them to 6-propagator integrals. We found that this step by itself 
is relatively straightforward and not too time-consuming; 
however, once the reduction of the produced 6-propagator integrals is attempted, the  
reduction procedure stalls. In order to simplify this step as much as possible, we 
perform it only at the level of combinations of integrals that are actually 
required for computing the amplitude or  the differential equations for the  master 
integrals, but not for all contributing integrals individually. 
Moreover, we want to work with as compact expressions as possible 
and we do this by choosing wisely the basis of the 7-propagator master
integrals. The main criterion that we impose is that, upon reduction, all 7-propagator integrals
are written in terms of master integrals, whose coefficients do not contain any non-factorizable 
unphysical poles which mix the Mandelstam variables ($s,t,u$) and the space-time
dimensions $d$.  We searched for the right basis empirically, by fixing the Mandelstam variables 
$s,t$ and $u$ to numerical values and performing the reductions with our code.\footnote{
This step can be equally well performed  using  any public reduction code.}  
On one hand, using numerical values for the Mandelstam variables 
makes the reduction very fast; on the other hand, it does not 
change  the dependence of the final result on the space-time dimension. 
We found many different bases  which fulfill the $d$-factorization requirement 
and we have chosen the basis that leads to the  differential equations with the  
nicest properties, as explained  in detail in Section~\ref{sec:deqs}.  The steps described 
above allowed us to express all the integrals that contribute to the scattering amplitudes 
in terms of master integrals and to derive the differential equations for master integrals 
retaining full dependence on the internal quark mass.  We will now explain how these
 differential equations 
are  used to compute the master integrals.  

\begin{figure}[t!]
\centering
\includegraphics[width=0.30 \linewidth]{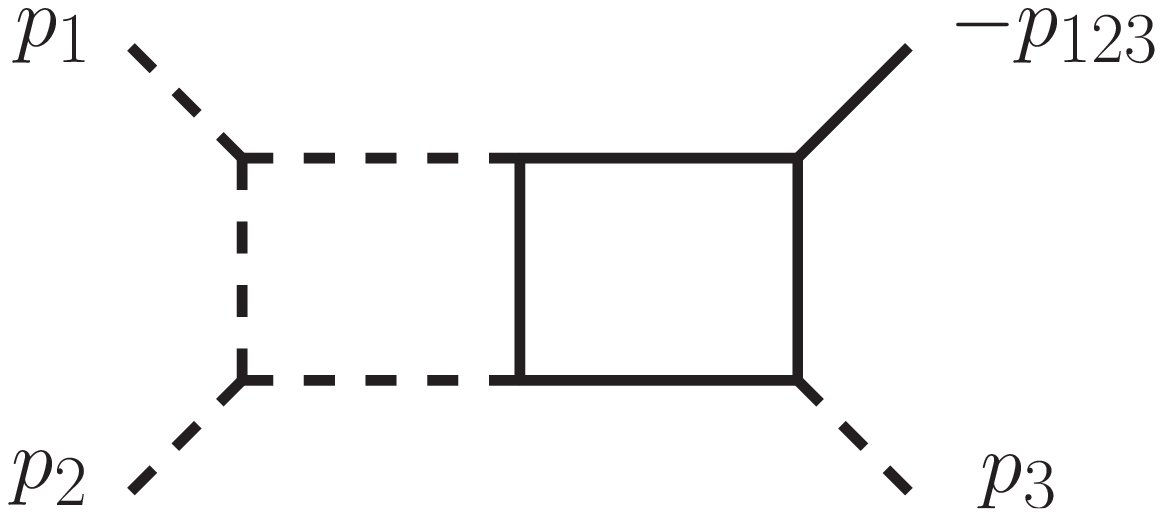}  \hspace{0.4 cm}
\includegraphics[width=0.30 \linewidth]{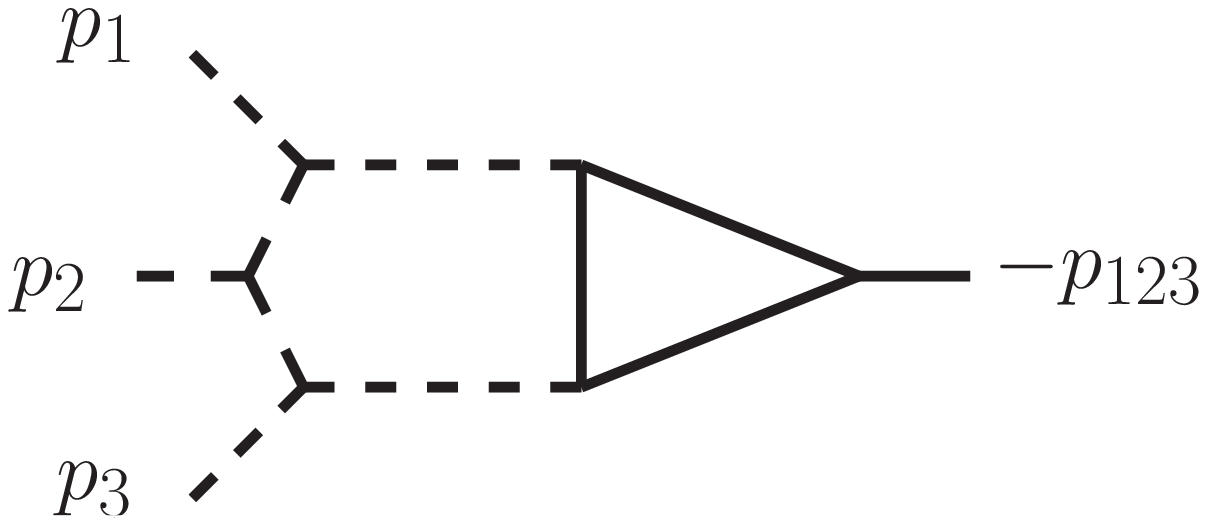} \\
\includegraphics[width=0.30 \linewidth]{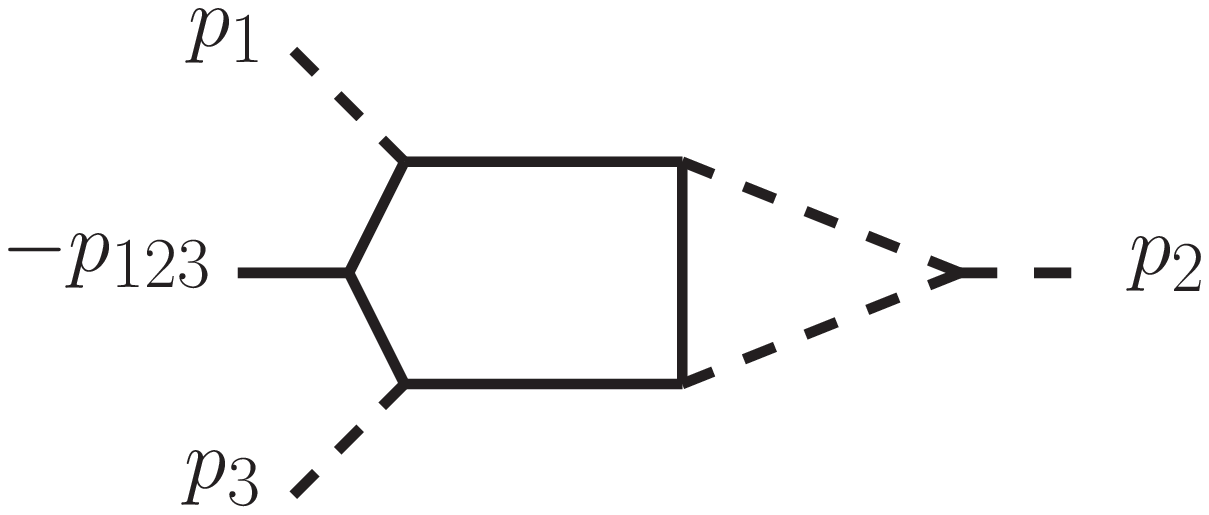}  \hspace{0.4 cm}
\includegraphics[width=0.30 \linewidth]{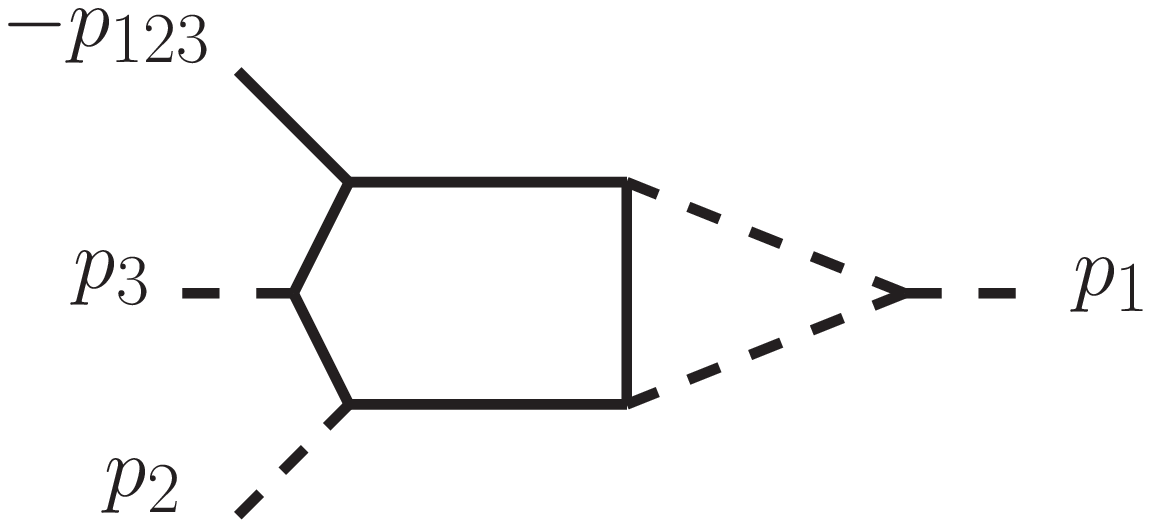} \hspace{0.4 cm}
\caption{Seven-propagator Feynman integrals  of the PL1 family,  
that appear in the form factors and require IBP reduction. 
The two integrals at the top are irreducible and correspond to two sectors in 
the family PL1 that contain master integrals with seven propagators. 
The integrals at the bottom  correspond to reducible integrals. All momenta are incoming.}  
\label{fig::PL1}
\end{figure}  

\begin{figure}[t!]
\centering
\includegraphics[width=0.30 \linewidth]{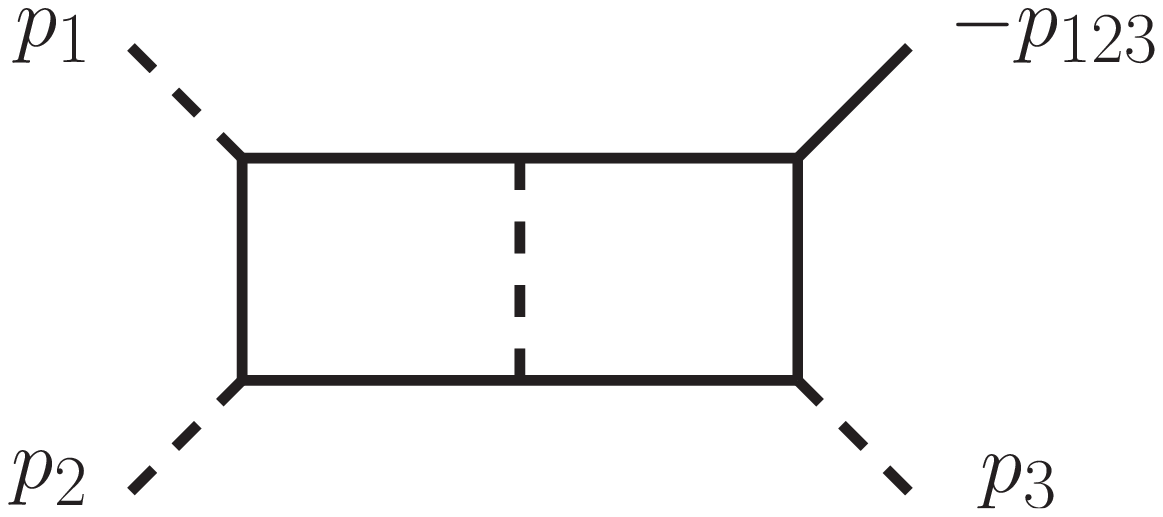} \\
\includegraphics[width=0.30 \linewidth]{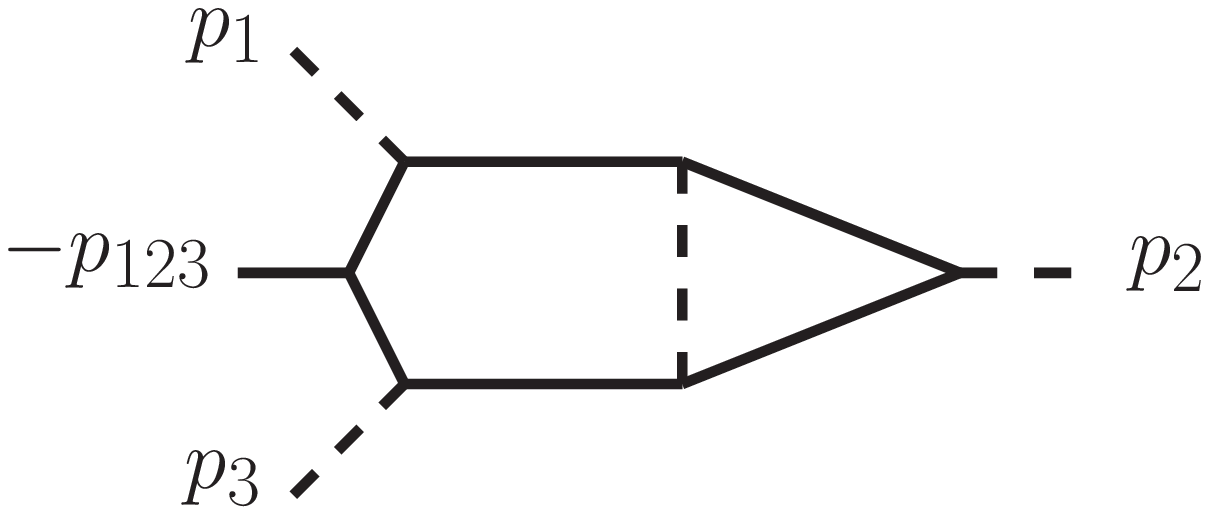} \hspace{0.4 cm}
\includegraphics[width=0.30 \linewidth]{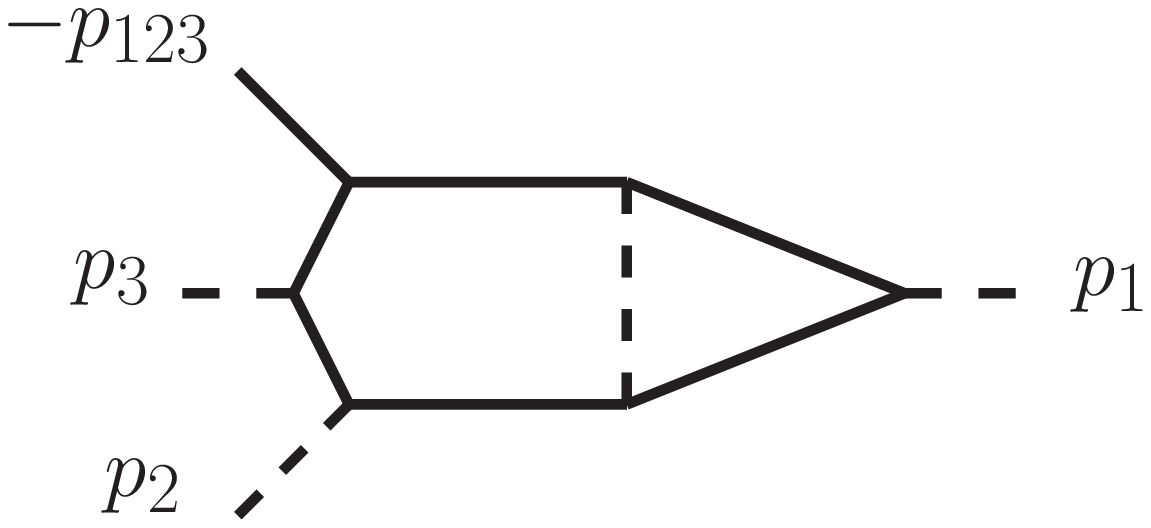} \hspace{0.4 cm}
\includegraphics[width=0.30 \linewidth]{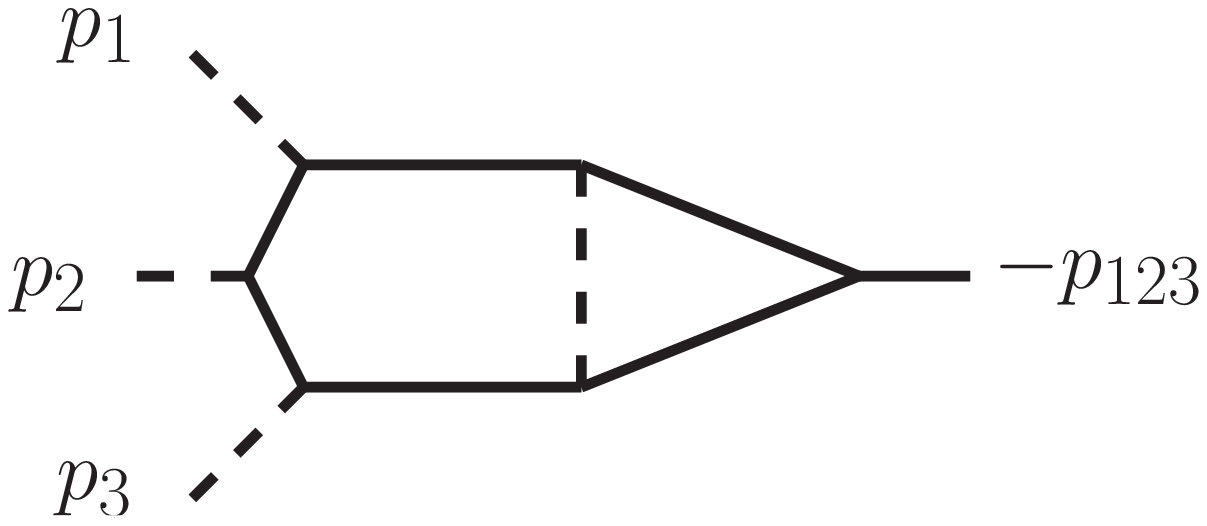} 
\caption{Same as in Figure \ref{fig::PL1}, but for the integrals of the PL2 family.}
\label{fig::PL2}
\end{figure}  

\begin{figure}[t!]
\centering
\includegraphics[width=0.30 \linewidth]{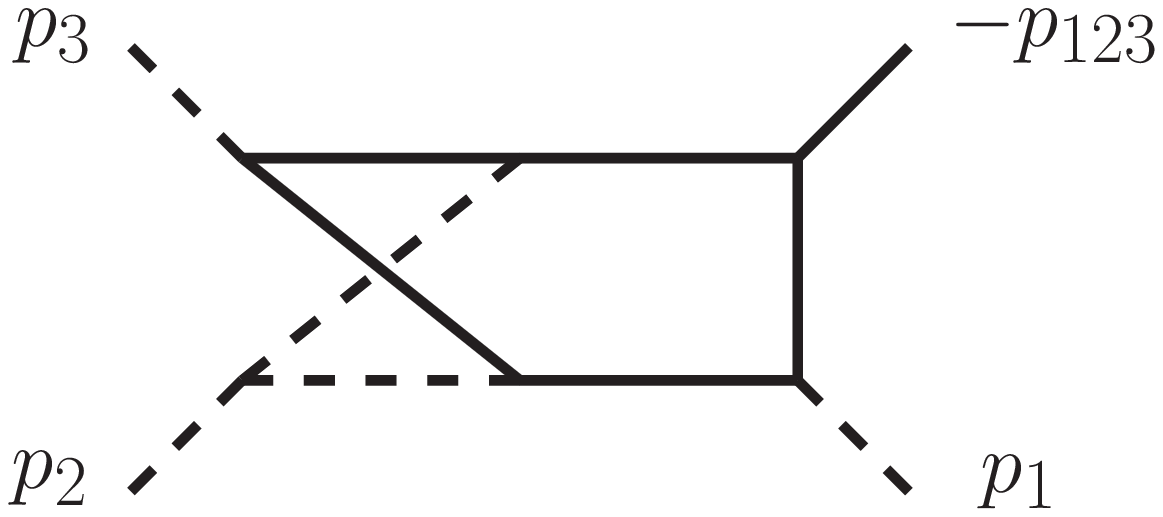} \hspace{0.4 cm}
\includegraphics[width=0.30 \linewidth]{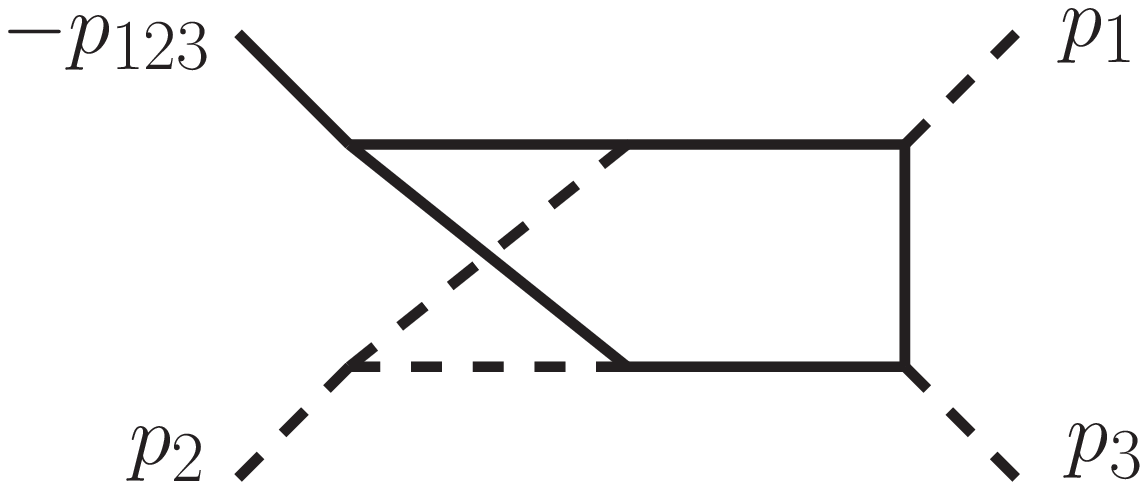} \hspace{0.4 cm}
\includegraphics[width=0.30 \linewidth]{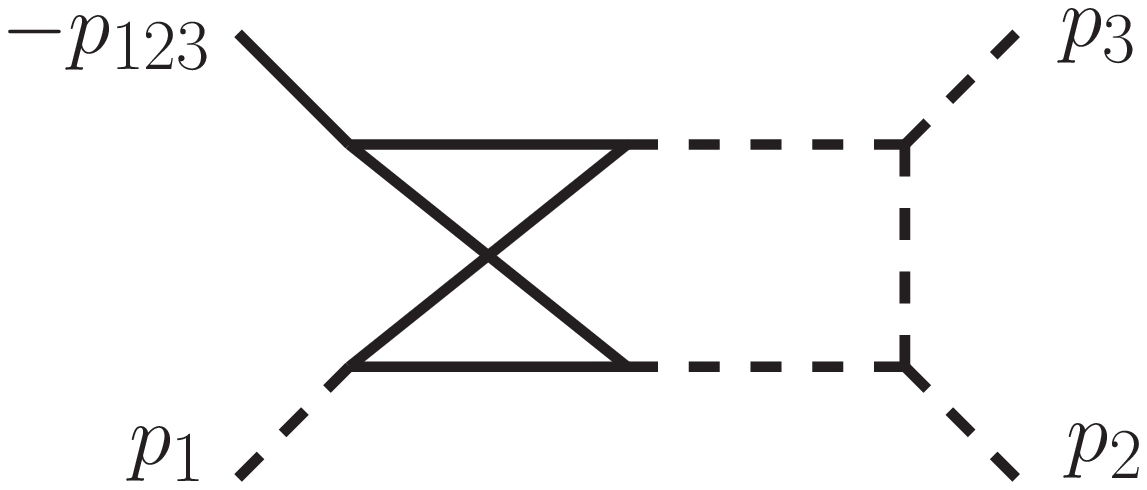}
\caption{Same as  in Figure \ref{fig::PL1} but for the integrals of the  NPL family. 
Note that in this case all the three integrals are irreducible. 
The leftmost  integral does not contribute to  the form factors because of the color 
structure of the corresponding Feynman diagrams. For this reason we do not compute it. 
}  
\label{fig::NPL}
\end{figure}  

\section{Solving the differential equations for the master integrals}\label{sec:deqs}

Following the procedure outlined in the previous Section, we write the form factors 
as linear combinations of the master integrals. Examples of master integrals with seven 
propagators that need to be computed  are shown  in Figs.~\ref{fig::PL1}, \ref{fig::PL2},  \ref{fig::NPL}.
Master integrals with six or less propagators are obtained from the ones with seven  
by removing  some of the internal lines.

To compute these master integrals, we consider their derivatives with respect to the  kinematic 
variables that they depend upon. These derivatives  are given by Feynman integrals that 
belong to the integral families that we discussed in the previous Section; for this reason  
they can be expressed   through the master integrals. Following these steps, we obtain a system 
of partial differential equations that the master integrals satisfy.

Derivation of the differential equations is 
facilitated by the fact that derivatives with respect to  kinematic 
invariants  can be written as linear combinations of derivatives with respect to the four-momenta 
of external particles; the latter  derivatives can be easily computed if we use Eq.(\ref{eq3.5a})  
to write down the master integrals. 
Specifically, treating $s$, $t$ and $u$ as independent variables, we obtain
\begin{eqnarray}
s\, \partial_{s}&=&\frac{1}{2}\left(p_1 \cdot\partial_{p_1}+p_2\cdot\partial_{p_2}-p_3\cdot\partial_{p_3}\right), \nonumber\\
t\, \partial_{t}&=&\frac{1}{2}\left(p_1\cdot\partial_{p_1}-p_2\cdot\partial_{p_2}+p_3\cdot\partial_{p_3}\right), \label{eq:deqsij}\\
u\, \partial_{u}&=&\frac{1}{2}\left(-p_1\cdot\partial_{p_1}+p_2\cdot\partial_{p_2}+p_3\cdot\partial_{p_3}\right), \nonumber
\end{eqnarray}
where $p_i \cdot \partial_{p_j} = p_i^\mu \, \partial/ \partial p_j^\mu$.
Partial derivatives with respect to  $y = t/m_h^2$ and $z = u/m_h^2$ {\it at fixed } $m_h^2$
are then related to the partial derivatives in Eq.(\ref{eq:deqsij}) in a straightforward 
manner
\begin{gather}
\partial_y=m_h^2\left(\partial_{t}-\partial_{s}\right), \quad \partial_z=m_h^2\left(\partial_{u}-\partial_{s}\right), 
\label{eq:deqxyz}
\end{gather}
The partial derivative with respect to the the $b$-quark mass is trivially related to 
the $\kappa$-derivative, $\partial_\kappa = -m_h^2 \partial_{m_b^2}$.

The differential equations are computed by applying  Eq.(\ref{eq:deqxyz})  
and Eq.\eqref{eq:deqsij} 
to  the master integrals and  using the integration-by-parts identities to reduce  
the resulting integrals    to master integrals. 
In this way, coupled systems of differential equations in $\kappa,\;y$ and $z$ are found for the  
list of master integrals that we denote  by $\{\I_i\}$ throughout this section.

We can also compute the derivatives of the master integrals with 
respect to $m_h^2$. However, 
these differential equations are not useful since, when the integrals
are expressed in terms of the variables $\kappa,y,z$ and $m_h^2$, the $m_h^2$-differential 
equations trivialize and only provide the (already known) information on the 
 canonical mass dimensions of the master integrals. 
For this reason,  we set $m_h^2=1$ at the beginning and 
re-introduce it back at the very end of the computation. 

The differential equations in $\kappa,y$ and $z$ take the following form 
\begin{equation}
\partial_k\I_i(\kappa,y,z,\epsilon)=A^k_{ij}(\kappa,y,z,\epsilon)\, \I_j(\kappa,y,z,\epsilon), 
\quad k \in \{ \kappa,\, y,\, z \}. \label{eq:deqm2yz}
\end{equation}
Matrices $A^k$ are rational functions of  $\kappa,y,z$ and $\epsilon$.  It is essential  that 
these matrices are sparse and, to a large extent,  triangular. This allows us to 
organize the process of solving the differential equations by starting from the simplest 
integrals with smaller number of propagators and gradually moving to more complex ones. 
The two-loop tadpole integral is  computed independently and used 
as an input for the differential equations for master integrals with three or more propagators. 

We are interested in solving the differential equations as an expansion in 
the normalized $b$-quark mass  squared $\kappa$,  
around $\kappa = 0$. From the structure of the 
differential equations it follows that  $\kappa = 0$ is a singular 
point; as the consequence, we have to look  for the solutions 
of the differential equations using the following 
Ansatz 
\begin{equation}
\I_i(\kappa,y,z,\epsilon)=\sum_{j,k\in\mathds{Z},n\in\mathds{N}} c_{i,j,k,n}(y,z,\epsilon)\, \kappa^{j-k\epsilon}\, \log^n(\kappa). \label{eq:ansatz1}
\end{equation}

In practice we observed 
that for the computation of the master integrals that appear in the  form factors, a  simpler Ansatz is 
sufficient\footnote{Some sectors 
in the non-planar family $\text{NPL}$ contain master integrals that 
scale like $( \kappa )^{-1/2}$ 
after expanding in $\kappa$, namely the two sectors $(1,1,0,1,0,1,1,1,0)$ and $(1,1,1,1,0,1,1,1,0)$. However the 
integrals in these sectors do not appear in the final reduced amplitude since their 
color factors vanish.}
\begin{equation}
\I_i(\kappa,y,z,\epsilon)=\sum_{j\geq-1}\sum_{k=0}^2\sum_{n=0}^2 c_{i,j,k,n}(y,z,\epsilon)\, 
\kappa^{j-k\epsilon}\, 
\log^n(\kappa). \label{eq:ansatz2}
\end{equation}

As indicated in Eq.(\ref{eq:ansatz2}), the strongest $\kappa$-singularity that we encountered in 
the master 
integrals is $\kappa^{-1}$; this is related to the fact that the  master integrals that we have chosen 
have at most one propagator raised to the second power.  As a rule, integrals with higher powers of propagators 
have stronger $\kappa$-singularities.  In principle, since we are interested in the computation of the  scattering amplitude 
in the limit $\kappa \to 0$ and since the scattering amplitude has at most logarithmic 
$\kappa$-singularities in this limit, 
it would have seemed natural to choose master integrals with similar or weaker singularities.  
 However, if this is done, it becomes  more  difficult to solve the system of differential equations. 
This can be understood 
by a closer proximity of integrals that have similar singularity structure 
in $\kappa$, compared to integrals whose  structure  of singularities is very different.

The coefficients $c_{i,j,k,n}$ are functions of $y,z,\epsilon$.  To determine  them, we start with 
 the differential equations 
with respect to $\kappa$.  We use  the Ansatz Eq.~\eqref{eq:ansatz2} 
in $\kappa$-differential equations 
and require that  the coefficients of the $\kappa^{j-k\epsilon} \log^n \kappa$  terms  
vanish independently of 
each other.  This gives a system 
of linear algebraic equations  for  the coefficients $c_{i,j,k,n}$ that can be solved 
straightforwardly.  

Suppose that a particular sector has $N$ coupled  master integrals. 
Upon solving the $\kappa$-differential 
equations in this sector,  we are left with $N$ unknown integration ``constants'' that are, in fact, 
functions of $y$ and $z$.  If in the massless case\footnote{Note 
that  by the ``massless 
case''  we mean  the limit 
$\kappa \to 0$ at fixed $\epsilon$.} there are $N_0$ master integrals in this 
 sector, then there are $N_0$  integration 
constants  that can be determined by matching the massive results to  the massless ones.  The massless  limit 
of the integrals that we study in this paper was 
computed in Refs.~\cite{Gehrmann:2000zt,Gehrmann:2001ck} and can be borrowed from there. 

The remaining $N-N_0$ coefficient functions have to be 
determined by considering the differential equations 
in $y$ and $z$. We use  the Ansatz Eq.~\eqref{eq:ansatz2} in the $(y,z)$-differential equations 
and again demand that the coefficients of the $\kappa^{j-k\epsilon} \log^n \kappa$  terms
vanish; this gives the required  $N-N_0$  $(y,z)$-differential equations for the coefficient functions. 
We solve these differential equations order by order in $\ep$.  We find that, similar to the massless 
case~\cite{Gehrmann:2000zt,Gehrmann:2001ck},  
the coefficient functions  can be expressed 
in terms of Goncharov  polylogarithms ({\sf GPL}). The {\sf GPL}'s are defined through 
the iterative formula 
\begin{gather}
\text{G}(\underbrace{l_1,\cdots ,l_n}_{\text{weight n}};x):=\int_0^x dx' \frac{\text{G}(l_2,\cdots ,l_n;x')}{x'-l_1},   \label{eq:2dHPL}
\end{gather}
subject to additional constraints 
\be
 \quad \text{G}(;x)=1, \ \ \ \ \text{G}(\underbrace{0,\cdots ,0}_{\text{n times}};x)=\frac{1}{n!}\log^n(x). 
\ee

The denominators that appear in the recursive integrands of the {\sf GPL}'s in Eq.(\ref{eq:2dHPL})
are the ones that appear in the matrices $A^k$ after expanding around
$\kappa = 0$; they assume the following values 
\begin{equation}
\{1-y-z,\, y,\, z,\, y+z,\, y-1,\, z-1\}. \label{eq:letters}
\end{equation}
It is easy to see from the definitions 
of {\sf GPL}'s that these denominators lead to  branch points at $x = 1-y-z$, $y,z=0$, $y=1$, $z =1 $ and $y = -z$. Physically, 
only the first three singular points are allowed while the last three are not. This implies that the corresponding 
{\sf GPL}'s can appear in the results for the master integrals 
only in such combinations where these unphysical singularities cancel. 
As we explain below, this feature allows us to simplify calculation of master integrals in certain cases. 

We  expand  the coefficient functions $c_{i,j,k,n}$   in $\epsilon$ through 
the  weight four.\footnote{For those branches where the expansion in $\epsilon$ results in only rational functions in $y$ and $z$, we expanded the solution of $c_{i,j,k,n}$ to exactly four orders higher in $\epsilon$, starting from the highest pole in $\epsilon$ of that branch.}  
We also adjust the expansion of the master integrals   in $\kappa$ in such a way  that the leading 
${\cal O}(\kappa)$ contribution 
to the amplitude can  be computed. Note that this requires expanding some of the integrals to relatively high order 
in $\kappa$ since some master integrals appear in the differential equations with coefficients 
that scale as $\kappa^{-n}$, $n > 0$,  in the $\kappa \to 0$ limit. 

Upon solving the differential equations, we write the solutions for each of the master integrals 
in the following way 
\begin{equation}
\I_i(\kappa,y,z,\epsilon)=\sum_{j=-1}^{j_{max}^{(i)}}\sum_{k=0}^2\sum_{n=0}^2 \, \, \sum_{r=r_0^{(i,j,k,n)}}^{r_0^{(i,j,k,n)}+4}\epsilon^r\, c_{i,j,k,n}^{(r)}(y,z)\, \kappa^{j-k\epsilon}\, \log^n(\kappa). \label{eq:mastersol}
\end{equation}
The lower limit of the $\epsilon$ expansion, $r_0^{(i,j,k,n)}$, is bounded below by $-4$. We include ancillary files with 
the paper that contain all master integrals required for computing the 
form factors, expanded in $\kappa$ and $\epsilon$ as in Eq.~\eqref{eq:mastersol}.  Note that this form of the solution gives 
access to different $\kappa$-branches, i.e. terms ${\cal O}(\kappa^{-2\ep},\kappa^{-\ep}, \kappa^0)$. Each of these branches 
should correspond to contributions of distinct  ``regions'', using the language of the 
``strategy-of-regions''  ~\cite{Beneke:1997zp,Smirnov:2002pj},  or ``modes'', in the language of effective field theories. 
Knowing the results for  each of the $\kappa$-branches 
separately should  be useful for understanding how to resum 
the $\log (\kappa)$ terms. Note also that individual branches have stronger $\ep$-singularities than 
the complete integral. 

Finally, we note that after solving the $(y,z)$-differential equations, we can only 
determine the master integrals up to the constants of integration, 
that need to be computed separately. 
We explain how to do this in the next Section. However, once this is done, we have the 
complete expression for the master integral and we can 
check it by  comparing numerically the expansions in $\kappa$ and $\epsilon$ of all 
the master integrals  in various kinematic  points in the Euclidean region with 
FIESTA~\cite{Smirnov:2015mct}.
For all integrals required for the calculation of the $gg \to Hg$ amplitude,
we found a perfect agreement between the analytical results obtained in this
paper and the numerical results obtained with FIESTA.

\section{Boundary conditions}
\label{sec:bound}

By solving the differential equations, we can only obtain  the master 
integrals up to the  integration constants.  These constants 
have to be determined separately. To this end, it suffices to 
compute the   master integrals at an arbitrary  
kinematic point and then match the results to the solutions 
of the differential equations. However, this procedure is our last resort since, 
usually, there are other, simpler, ways to obtain 
the required integration constants.

We have already mentioned some of them in the previous Section. 
For example, if we are able to determine a master integral 
from the $\kappa$-differential equation,
the integration constant is the massless branch of a particular integral known 
from Refs.~\cite{Gehrmann:2000zt,Gehrmann:2001ck}. 

Another way  to determine the integration constants arises if 
the homogeneous parts of  the $(y,z)$-differential equations  
exhibit unphysical singularities in $x,y$ and $z$ variables. Since, upon 
integration, singularities of  differential equations become 
singularities of  master integrals and since only certain, physical, singularities  
in $x$, $y$ and $z$ can appear in master integrals, we  determine some of the integration 
constants by requiring that the
 unphysical singularities of the differential  equations do not appear in 
the master integrals. 

When none of the above applies, the integration constants have to be 
computed by matching  the value of an integral to the solution of the 
differential equations for some values of $y$ and $z$. 
It is difficult to describe how this has been  done since 
there is no  method that covers all the  cases. In practice,   we have used 
different techniques such as 
integration over  Feynman parameters,  expansion-by-regions, Mellin-Barnes integration 
and fitting numerical values of the integrals to linear combinations of 
the fixed-weight  irrational numbers, 
to determine integration constants for  master integrals. 
We give a few examples below   to illustrate how 
these techniques are applied. 

Consider the integral $\I_{\rm PL2}(1,1,0,0,1,0,1,0,1)$.  It is given by 
the following expression 
\be
\begin{split}
\I_{\rm PL2}(1,1,0,0,1,0,1,0,1) = & 
\int 
\frac{{\cal D} k_1 {\cal D} k_2 }{ ( (k_1 -p_2)^2 - m_b^2 ) ( (k_1 + p_1)^2 - m_b^2)} 
\\ 
& \times \frac{1}{(k_1 - k_2)^2 ( k_2^2 - m_b^2) ( (k_2 + p_1)^2 - m_b^2) 
}.
\label{eq4.1}
\end{split}
\ee
In the limit of 
small $\kappa = - m_b^2/m_h^2$, the integral behaves as 
\be
m_h^2 \I_{\rm PL2}(1,1,0,0,1,0,1,0,1) = x^{-1} \left ( 
\kappa^{-2\ep} C_{2}
+\kappa^{-\ep} x^{-\ep} C_{1} 
+ {\cal O}(\kappa^0)
\right ).
\label{eq4.2}
\ee
The singularity at $x = 1-y-z = 0$ is  allowed and the two constants 
of integration $C_{1,2}$ can not be determined from the differential 
equations. To find these constants, we need to extract  the non-analytic 
terms that arise in the limit $\kappa \to 0$. 

We do this by first re-writing the integral over the loop momenta 
through an integral over Feynman parameters. We obtain 
\be
m_h^2\, \I_{\rm PL2}(1,1,0,0,1,0,1,0,1) = 
\frac{\Gamma(1+2\ep)}{\Gamma(1+\ep)^2}
\int \limits_{0}^{1} 
\frac{{\rm d} \alpha\, {\rm d} \beta\, {\rm d} \xi\, {\rm d} \mu \,
\beta^{-\ep} (1-\beta)^{\ep} \mu^{\ep} (1-\xi)^{1+\ep}
}{\Delta^{1+2\ep} },
\ee
where 
\be
\Delta =  x ( 1 - \mu (1-\alpha) ) \xi(1-\xi) (1-\beta) 
+ \kappa  ( 1 - \beta (1 - \mu(1 - \xi) ) )\,.
\ee
Inspecting the above equations, we conclude that the two  branches,  $\kappa^{-\ep}$ 
and $\kappa^{-2\ep}$,  appear due to the integration over two  distinct regions 
\be
\begin{split} 
& {\rm Branch}~\kappa^{-2 \ep}\;\;\;  \leftrightarrow \;\;\;  \xi \sim \kappa,\;\;
\alpha \sim \beta \sim \mu \sim 1,
\\
&  {\rm Branch}~ \kappa^{-\ep} \;\;\;  \leftrightarrow \;\;\; 1-\beta \sim \kappa,\;\;
\alpha \sim \xi \sim \mu \sim 1.
\end{split} 
\ee
To project the integrand on one of the two branches, one should 
Taylor expand the integrand in a variable that is small for a particular 
branch  and extend 
the integration over this variable to the positive half of the real axis. 
Upon   applying this 
procedure,  we arrive at the  following expression for the branch $\kappa^{-2\ep}$  and for the constant $C_2$
\be
C_2\; x^{-1} \kappa^{-2\ep} = 
\frac{\Gamma(1+2\ep)}{\Gamma(1+\ep)^2}
\int \limits_{0}^{1} 
{\rm d} \alpha\,  {\rm d} \beta\,  {\rm d} \mu \,
\beta^{-\ep} (1-\beta)^{\ep} \mu^{\ep} 
\int \limits_{0}^{\infty} 
 \frac{ {\rm d} \xi }{{\tilde \Delta}^{1+2\ep} },
\ee
where 
\be 
{\tilde \Delta}^{1+2\ep} = x ( 1 - \mu (1-\alpha) )(1-\beta) \xi 
+ \kappa ( 1 - \beta (1-\mu) ).
\ee
The integration over $\xi$ and $\alpha$ can be easily performed and we obtain 
\be
C_2 = 
 \frac{\Gamma(1+2\ep)}{2 \ep \Gamma(1+\ep)^2}
\int \limits_{0}^{1} \frac{{\rm d} \mu}{\mu^{1-\ep}} \ln ( 1 - \mu) 
\int \limits_{0}^{1} {\rm d} \beta
\frac{ \beta^{-\ep} (1-\beta)^{\ep - 1}}{ ( 1 - \beta(1-\mu) )^{2\ep} }.
\ee
Upon writing 
\be
(1-\beta(1-\mu) )^{-2\ep} = \mu^{-2\ep} + 
\left [ (1-\beta(1-\mu) )^{-2\ep} - \mu^{-2\ep} \right], 
\ee
we split the integral into two parts; the first integral 
can be evaluated exactly and the second can be evaluated 
upon expanding  in $\epsilon$.  We find 
\be
C_2  = 
-\frac{\pi^2}{12 \ep^2} - \frac{\zeta_3}{2 \ep}  - \frac{\pi^4}{72}.
\ee

To obtain the second constant, we need to understand how the 
 branch ${\cal O}(\kappa^{-\ep})$ arises in the  integral. As we already 
pointed out, this branch corresponds to the region where $1-y \sim \kappa$. 
We therefore change variables $y \to 1-f$, Taylor expand the integrand 
in $f$ assuming that $f \sim \kappa \ll \alpha, \mu, \xi$ and extend the integration 
over $f$ to  the positive half of the real axis. We find 
\be
C_1 \; x^{-1-\ep}  \kappa^{-\ep} =
\frac{\Gamma(1+2\ep)}{\Gamma(1+\ep)^2}
\int \limits_{0}^{\infty}  {\rm d} f
\int \limits_{0}^{1} 
 \frac{ 
{\rm d} \alpha\, {\rm d} \xi \, {\rm d} \mu \;
\mu^{\ep} (1-\xi)^{-\ep}  f^{\ep}  }{
\left ( x (1- \mu (1-\alpha) ) \xi f + \kappa \mu \right )^{1+2\ep} },
\ee
Integrations over $f, \xi$ and $\mu$ are straightforward and 
we obtain 
\be
\begin{split} 
C_1  =   
\frac{ \Gamma(1-\ep)^2}{\ep^3 \Gamma(1-2\ep)} 
\lim_{\nu \to 0} \left [ \frac{1}{\nu} - B(\nu,1-\ep)  \right 
] = 
\left ( -\frac{\pi^2}{6 \ep^2} - \frac{\zeta_3}{\ep} + \frac{\pi^4}{60}
\right ).
\end{split} 
\ee

As the second, more complicated  example, we consider the evaluation 
of one of the non-planar, seven-propagator integrals. The integral 
reads
\be
\begin{split} 
& \I_{\rm NPL}(2,1,1,1,0,1,1,1,0) =   
\int 
\frac{{\cal D} k \; {\cal D} l }{ ( k^2 - m_b^2)^2 
 ( (k+p_2)^2 - m_b^2) ( (k - p_3)^2 - m_b^2)  } \\
&\;\;\;\;\;\;\;\;\;\;\;\; \times \frac{1}{
  ((k-l-p_1 - p_3)^2 - m_b^2)
  ( (k-l+p_2)^2 - m_b^2 ) 
l^2  ( l+p_1)^2
}.
\label{eq4.13}
\end{split}
\ee
The constants of integration that need to be determined 
are in the  branch $\kappa^{-\ep}$, 
including the coefficient of the  
logarithmic term $\kappa^{-\ep} \log \kappa $. 
To determine these constants, we compute the integral at  a particular 
kinematic point $z \to 1, y \to 1$. This implies that $x = 1- y -z = -1$, 
so that the integral receives an imaginary part. 

To compute the integral Eq.(\ref{eq4.13}) at the kinematic point described 
above in the limit $\kappa \to 0$, we write  it as an 
integral over suitably chosen  Feynman parameters. In particular, 
we start by combining two pairs of propagators into single propagators
\be
\begin{split} 
& \frac{1}{l^2(l+p_1)^2} = \int \limits_{0}^{1} 
\frac{{\rm d} \eta}{ ( ( l+ p_1 \eta)^2 )^2 },
\\
& \frac{1}{(k^2 - m_b^2)^2 ( (k+p_2)^2 - m_b^2) } = 
2 \int \limits_{0}^{1} 
\frac{{\rm d} \xi \;  (1-\xi) }{ ( ( k+ p_2 \xi)^2 - m_b^2 )^3 },
\end{split} 
\ee
and then shift the loop momenta $ l \to l + k - \eta p_1$ 
and $ k \to k + (1-\eta) p_1 + p_3 $. The integral becomes  
\be
\begin{split}
& \I_{\rm NPL}(2,1,1,1,0,1,1,1,0)  = 2\int \limits_{0}^{1} 
{\rm d} \eta {\rm d} \xi (1 - \xi) 
\int 
\frac{ {\cal D} k \, {\cal D}l }{ 
( ( k + p^\eta_1 )^2 - m_b^2)
}
\\
& \times \frac{1}{ ( ( k + p^\eta_1 + p_3 + p_2 \xi)^2 - m_b^2)^3  (l^2 - m_b^2) ( (l - p_H)^2 - m_b^2) (l+k)^2 },
\end{split} 
\ee
where $p_1^\eta = (1-\eta) p_1$ and $p_H = -p_1 -p_2 - p_3$.
Next, we integrate over $l$ and $k$, 
substitute $z = 1$, $y = 1$ and arrive at the following 
representation for the integral
\be
\begin{split}
& m_h^8 \I_{\rm NPL}(2,1,1,1,0,1,1,1,0) = \frac{\Gamma(4+2\ep) }{\Gamma(1+\ep)^2}
\int 
\frac{ {\rm d} \eta\, {\rm d} \xi \,{\rm d} x_3\, {\rm d} f \,
{\rm d} \rho\, {\rm d} t }{ ( 
\Delta + \kappa (t + (1-t) x_3) )^{4+2 \ep} 
}
\; 
\\
&\;\;\;\;\;\;\;\;\;\;  \times  (1-\xi) \;x_3^{3+\ep}  (1-x_3)^{-\ep-1} \;
\rho^2 \; t^{1+\ep} \;  (1-t)^3,
\label{eq4.16}
\end{split} 
\ee 
where 
\be
\Delta = \eta \rho (1-t)t x_3(1-\xi) + \rho(1-t)(1-\rho (1-t) ) x_3 \xi
+(1-f) f t + f(f-2 \rho) (1-t) t x_3.
\ee
We notice that the integrand is linear in $\eta$ and $\xi$. We also 
notice that the coefficient of ${\cal O}(\eta)$ term in $\Delta$ is 
proportional to $1-\xi$; this means  that upon integration over $\eta$, it cancels 
the  $(1-\xi)$ factor in the integrand Eq.(\ref{eq4.16}). As the result,  the 
 integrations over $\eta$ and $\xi$ can be performed exactly.
Integrating over $\eta$ and $\xi$, we arrive at the following 
represention of the non-planar integral at the kinematic point $z = 1, y = 1$
\be
\begin{split} 
& m_h^8 \I_{\rm NPL}(2,1,1,1,0,1,1,1,0)
 = \frac{\Gamma(2+2\ep)}{\Gamma(1+\ep)^2} 
\left ( 
\tilde \I_a + \tilde \I_b  \right ),
\\ 
& \tilde \I_a  = 
\int {\rm d} x_3\, {\rm d} f\, {\rm d} \rho\, {\rm d} t 
\frac{ x_3^{1+\ep} (1-x_3)^{-\ep - 1} t^{\ep} (1-t)}{ 
 ( 1 - \rho(1-t) )  }
\left [ 
\Delta_{00}^{-2-2\ep} - \Delta_{01}^{-2-2\ep}
\right ],
\\
& \tilde \I_b  = 
-\int {\rm d} x_3\, {\rm d} f\, {\rm d} \rho\, {\rm d} t \,
\frac{ x_3^{1+\ep} (1-x_3)^{-\ep - 1} t^{\ep} }{ 
 ( 1 - \rho  ) }
\left [ 
\Delta_{10}^{-2-2\ep} - \Delta_{11}^{-2-2\ep}
\right ],
\label{eq4.18}
\end{split} 
\ee
where 
\be
\begin{split} 
& \Delta_{00} = f(1-f) t + f(f-2 \rho) (1-t) t x_3 + \kappa( t + (1-t) x_3),
\\
& \Delta_{01} = f(1-f) t + (1-t) x_3 ( \rho(1-\rho) + (f - \rho)^2 t) 
+ \kappa (t + (1-t) x_3), 
\\
& \Delta_{10} = f(1-f) t + (1-t) t x_3 ( (f - \rho)^2 + \rho(1-\rho) ) 
+ \kappa ( t + (1-t) x_3), 
\\
& \Delta_{11} = f(1-f)  t + (1-t) x_3 ((f-\rho)^2 t +  \rho(1-\rho)   ) 
+ \kappa  ( t + (1-t) x_3).
\end{split} 
\ee

To determine the coefficient of  the $\kappa^{-\ep -1}$ branch,  that arises in the 
limit $\kappa \to 0$, we need to consider two integration regions
\be
{\rm Region~1}: t \sim \kappa,  x_3 \sim \rho \sim f \sim 1,
\;\;\; {\rm Region~2}: t \sim (1-\rho) \sim \kappa ,\;\; x_3 \sim f \sim 1.
\ee
As always, we perform the Taylor expansion of the 
integrand in all the variables that are parametrically small 
 and then extend the integration region over the small variables 
to  the positive half of the real axis. An important comment worth making 
is that, in order to be able to treat all the different terms 
in Eq.(\ref{eq4.18}) separately, one has to introduce an additional 
regulator; in particular, the prefactors $(1-\rho(1-t) )$ and $(1-\rho)$ in Eq.(\ref{eq4.18})
must be modified in the following way
\be
(1-\rho(1-t) ) \to (1-\rho(1-t) )^{1+\nu},\;\;\;\; (1-\rho) \to (1-\rho)^{1+\nu}.
\ee

Constructing the expansion of Eq.(\ref{eq4.18}) along the lines sketched 
above, we find that the $\kappa^{-\ep - 1}$ branch of 
$\lim_{\kappa \to 0} \I_{\rm NPL}(2,1,1,1,0,1,1,1,0)$
is obtained as the  sum of five integrals.  We write 
\be
m_h^8 \I^{\kappa,-\ep-1}_{\rm NPL}(2,1,1,1,0,1,1,1,0)
= \kappa^{-1-\ep} \lim_{\nu \to 0} \left ( 
I_1 + \kappa^{-\nu} I_2 + I_3 + I_4 +  \kappa^{-\nu} I_5 \right ).
\label{eq5.22}
\ee
All of these integrals involve integration over some 
``small'' variables that can be easily performed since 
the corresponding integration region extends from zero to infinity.
We present the expressions for the five integrals after these simple 
integrations are carried out. We find 
\be
\begin{split} 
& I_{1} =  
 \int \frac{{\rm d} x_3 {\rm d} f {\rm d}\rho 
(1-x_3)^{-\ep-1} f^{-1-\ep} }{(1-\rho)^{1+\nu} ( 1- f + x_3 (f-2\rho) )^{1+\ep}}, 
\\
& I_{2} =
\frac{\Gamma(1+\ep + \nu) \Gamma(1+\ep-\nu)}{
\nu \Gamma(1+\ep)^2} 
\int  \frac{ {\rm d} x_3 {\rm d} f \; (1-x_3)^{-\ep-1} x^{-\nu}}{
(f (1-f) +f(f-2) x_3 )^{1+\ep - \nu}
},
\\
& I_3 = -  
\int \limits_{0}^{\infty} {\rm d} \xi 
\int \limits_{0}^{1} 
\frac{ {\rm d}x_3 {\rm d} f  \; (1-x_3)^{-\ep-1}  (1-f)^{-\ep}
( x_3 + f(1-x_3) )^{-\ep}  }{ (1+\xi)^{1+\ep}
( x_3 + (1-f)(x_3 + f(1-x_3) ) \xi ) 
},
\\
& I_4 = -  
\int \frac{{\rm d} x_3 {\rm d} f {\rm d} \rho (1-x_3)^{-\ep-1} 
(1-\rho)^{-\nu -1 } }{ ( f (1-f) + ( (f-\rho)^2 + \rho(1-\rho) ) x_3)^{1+\ep} },
\\
& I_5 = 
\int \limits_{0}^{1} {\rm d} x_3 {\rm d} f \int \limits_{0}^{\infty} 
{\rm d}\bar \rho \frac{ (1-x_3)^{-\ep - 1} (1-f)^{-\ep-1} }{
{\bar \rho}^{1+\nu} ( \bar \rho + 1)^{1+\ep} 
\left ( f + x_3 (1-f) \right )^{1+\ep}
}.
\end{split} 
\ee
We compute these integrals directly.  We note that through weight three, 
it is straightforward to calculate them analytically.  However, 
at weight four, we compute some of the integrals numerically and 
then fit them to a linear combination of the irrational constants of weight four 
that include $\log 2$ to an appropriate power and 
$\Li_4(1/2)$.  Working through ${\cal O}(\nu^0, \ep^1)$,  we obtain\footnote{We only show 
the real parts of the integrals $I_{1,..,5}$.} 
\be
\begin{split} 
I_1 = & \frac{1}{\nu} \left ( 
 \frac{1}{\ep^2} - 
\frac{4 \pi^2}{3} + \ep \left (-5 \pi^2 \log 2 - \frac{13}{2} \zeta_3 \right ) \right ) 
- \frac{3 \pi^2}{2\ep} - \left ( \frac{5 \pi^2}{2} \log 2 + \frac{35}{4} \zeta_3 \right ) 
\\
& + \ep \left ( \frac{5 \pi^4}{144} - \frac{4  \pi^2}{3} \log^2 2 + \frac{5}{6} \log^4 2
+ 20 {\rm Li}_4(1/2) \right ),
\\
I_2 = & \frac{1}{\nu} \left ( -\frac{1}{\ep^2} + \frac{4 \pi^2}{3} + \ep \left ( \pi^2 \log 2 + \frac{13}{2} \zeta_3 \right ) \right )
-\frac{1}{\ep^3} - \frac{\pi^2}{3 \ep} + 2\zeta_3 
\\ &  + \ep \left ( \frac{37 \pi^4}{120} + 
 \frac{4 \pi^2 \log^2 2}{3}   - \frac{\log^4 2}{3}  - 8 {\rm Li}_4(1/2)  \right ),
\\
I_3 = & \frac{\pi^2}{6\ep} -3 \zeta_3 + \ep \left ( 
-\frac{37 \pi^4}{360} - \frac{2 \pi^2}{3} \log^2 2
+ \frac{2 \log^4 2 }{3} + 16 {\rm Li}_4(1/2)
\right ),
\\
I_4 = &  \frac{1}{\nu}
\left ( \frac{1}{\ep^2} + \frac{\pi^2}{6} + 4 \zeta_3 \ep \right )
-\frac{1}{\ep^3} + \frac{\pi^2}{2\ep} + \pi^2 \log 2 + \frac{3}{2} \zeta_3
\\
& + \ep \left ( \frac{11 \pi^4}{72} + \frac{2 \pi^2 \log^2 2}{3}  
- \frac{2}{3} \log^4 2 -  16 {\rm Li}_4(1/2) \right ),
\\
I_5 = & \frac{1}{\nu} \left ( -\frac{1}{\ep^2} - \frac{\pi^2}{6} - 4 \zeta_3 \ep \right ) 
- \frac{\pi^2}{6\ep} + \zeta_3  - \frac{7 \ep \pi^4}{180} .
\end{split} 
\ee
Using these results in Eq.(\ref{eq5.22}), we obtain the coefficient  of the 
$\kappa^{-1-\ep}$ branch  of the 
integral $\lim_{\kappa \to 0} \I_{\rm NPL}(2,1,1,1,0,1,1,1,0)$ that arises in the $
\kappa \to 0$ limit, 
at the point $z = 1, y=1$. This result is then 
matched to the solution  found for this branch from the $(y,z)$-differential equation 
and the constant of integration is determined.  Other branches of this integral, 
as well as 
other integrals, that require 
determination of the integration constants can be studied along 
similar lines.

\section{Helicity amplitudes} \label{sec:ancont}

In this Section, we present the results of the computation of
the  $H \to ggg$ scattering amplitude. 
It is convenient to write the result fixing gluon helicities.
We obtain  the helicity  amplitudes from  the general 
expression for the $H \to ggg$ amplitude given in Eqs.(\ref{eq:ampl},
\ref{eq:tensampl}).
We write 
\be
{\cal A}_{\lambda_1 \lambda_2 \lambda_3}(s,t,u) = 
\epsilon_{1,\lambda_1}^\mu (p_1) 
\epsilon_{2,\lambda_2}^\nu (p_2) \epsilon_{3,\lambda_3}^\rho (p_3)
{\cal A}_{\mu \nu \rho}(s,t,u), \label{eq:helamplprel}
\ee
where $\lambda_{1,2,3}$ 
are helicity labels  of gluons with momenta $p_{1,2,3}$ respectively, 
and the dependence of the amplitude on the $b$-quark mass is suppressed. 
Since each gluon  has two independent  polarizations, there 
are eight helicity amplitudes 
in total  but only two of them are independent. The 
remaining six amplitudes can be obtained 
by permuting the external gluons and applying parity and charge conjugation. 

We choose  the two amplitudes ${\cal A}_{++ \pm }(s,t,u)$ 
as independent and write them using the spinor-helicity notations.\footnote{
See Ref.~\cite{Dixon:1996wi} for a review.}
The polarization vectors for external outgoing gluons with momentum $p_j$ and 
the reference vector $q_j$ read
\begin{align}
\epsilon_{j}^{\mu, +}= \frac{ \langle q_j | \gamma^\mu | j] }{ \sqrt{2} \langle q_j j \rangle }, 
\;\;\;
\epsilon_{j}^{\mu, -}= - \frac{[ q_j | \gamma^{\mu} | j \rangle }{\sqrt{2}\,  [ q_j j ]}.
\end{align}
We note that the reference vectors $q_j$ must be chosen for each gluon  in accord with Eq.~\eqref{eq:gauge}.
Using  these notations, we write the two independent helicity amplitudes as  
\be
\begin{split}
&{\cal A}_{+++}(s,t,u,m_b) = 
\frac{m_h^2}{\sqrt{2} \langle 12 \rangle \langle 23 \rangle \langle 31 \rangle}\; 
\, \Omega_{+++}(s,t,u,m_b)\,,
\\
&{\cal A}_{++-}(s,t,u,m_b) = \frac{[12]^3}{\sqrt{2}\, [13]\, [23]\, m_h^2}\; 
\, \Omega_{++-}(s,t,u,m_b)\,.
\label{eq:helampl}
\end{split}
\ee
The helicity 
coefficients $\Omega_j$ are linear combinations of the form factors
defined in Section~\ref{sec:notation}; they read 
\be
\Omega_{+++} = 
\frac{su}{m_h^2}
\left( 
 F_1 
+ \frac{t}{u} F_2 + \frac{t}{s} F_3 + \frac{t}{2} F_4 
\right ),
\;\;\;\;\;
\Omega_{++-} = 
\frac{m_h^2 u}{s} \left ( F_1 + \frac{t}{2} F_4  \right ) \,. \label{eq:helcoeff}
\ee

Similar to the form factors, the helicity 
coefficients $\Omega_j$ can be written as an expansion in the 
strong coupling constant. We write 
\be
\Omega_{ ++ \pm } = \frac{m_b^2}{v}
\sqrt{ \frac{ \alpha_s^3  }{\pi} }
\left[  \;\; \Omega_{++ \pm}^{(1l)}  
 +  \left( \frac{\alpha_s}{2\,\pi}\right) \Omega_{++ \pm}^{(2l)} 
+ \mathcal{O}(\alpha_s^2) \;\;  \right   ]\,, 
\ee
where $m_b$ is the $b$-quark pole mass and $\alpha_s$ is the strong coupling constant 
at the scale $\mu = m_h$ defined in the  theory with $N_f$ active flavors. 
The two-loop helicity coefficients are not finite, but their 
infra-red divergent parts are described by the Catani's formula
\be
\Omega_{++\pm}^{(2l)} = I_1(\ep) \;  \Omega_{++\pm}^{(1l)} + \Omega_{++\pm}^{(2l),\fin},
\ee
where the $I_1(\ep)$ operator is defined  in Eq.(\ref{eq:cataniI1}).

We note that, originally, we defined the renormalized coupling 
constant in a scheme with $N_f$  active flavors since  the 
contribution of the $b$-quark loop was subtracted at zero momentum. 
This is not optimal since  we are interested in a kinematic situation 
where all momenta transfers are 
large compared to $m_b$.  In this case, the appropriate strong coupling to use is the one defined in the 
scheme with $N_f+1$ active flavors. The relation between the two couplings
at the scale $\mu = m_h$ reads 
\begin{equation}
\alpha_s^{(N_f)}  = \alpha_s^{(N_f+1)}
\left[ 1 - \frac{\alpha_s^{(N_f+1)} }{6 \pi} 
\log{\left( \frac{m_h^2}{m_b^2}\right)} + \mathcal{O}(\alpha_s^2) \right]. \label{eq:numflav}
\end{equation} 
We use this relation to re-write the helicity amplitudes using  the strong 
coupling constant defined in a theory with $N_f+1$ active flavors. 
Since the relation 
between the two coupling  constants is finite, it induces 
changes in the finite parts of the 
two-loop helicity amplitudes. 
We denote the helicity coefficients written with  the strong coupling 
in the theory with $N_f+1$ active flavors as $\overline \Omega$. 
We obtain 
\be
{\overline { \Omega}}_{++\pm}^{(1l),\fin} = \Omega_{++\pm}^{(1l),\fin},\,\;\;\;
{\overline \Omega }_{++\pm}^{(2l),\fin} = \Omega_{++\pm}^{(2l),\fin}
- \frac{1}{2} \log{\left( \frac{m_h^2}{m_b^2}\right)} {\overline \Omega}_{++\pm}^{(1l),\fin} \,.
\ee
We also note that it is far from obvious that the on-shell renormalization 
scheme  for the $b$-quark mass is, indeed, physically appropriate. The helicity
amplitudes are proportional to $m_b^2$, where one power comes from the 
Yukawa coupling constant and the other from the helicity flip in the quark 
loop.  It is most likely that the appropriate 
choice of the mass parameter related to the Yukawa 
coupling is the $\overline{\rm MS}$ mass defined at the scale $m_h$. However, 
the proper choice of 
the mass parameter responsible for the helicity flip is much less clear. 
It will be very interesting to understand the   scale and 
scheme  choices in the virtual 
amplitude that minimize the magnitude of the NLO QCD corrections 
to physical observables, for example to the Higgs transverse momentum 
distribution. We plan to return to this question in the future work. 

Full results for  helicity coefficients  ${\overline \Omega}_{++\pm}$  
can be found in  
the ancillary files provided with this submission. Although, 
on the scale of  known two-loop helicity amplitudes, 
our ${\overline \Omega}$'s are quite compact, 
they are nevertheless complex. It is therefore instructive  to study them in 
 a few interesting limits, where they simplify dramatically.

One such limit is the {\it soft limit}. It describes a situation where  
the energy of one of the gluons in the 
$H \to ggg $ amplitude becomes small. We take the gluon with momentum 
$p_3$  to be soft;
this implies the following hierarchical relations between 
the kinematic variables  
$m_b^2 \ll  t \sim u  \ll m_h^2 \sim s$. For the dimensionless variables 
introduced earlier, the soft limit  implies  $ \kappa \ll y \sim z \ll 1$.

It follows from Eq.~\eqref{eq:helampl} that  in the soft  limit 
the helicity amplitudes diverge as $ 1/\sqrt{y z}  \sim 1/\sqrt{p_\perp^2}$, 
where we introduced the transverse momentum of the Higgs boson 
$p_\perp^2 = u t /s$. This is the standard soft 
  divergence present in any scattering amplitude. 

The helicity coefficients ${\overline \Omega}_{++\pm}$ develop logarithmic singularities in the soft limit. 
It is convenient to write these helicity coefficients in terms of 
the logarithms of the ratio of the bottom quark 
mass and the Higgs boson mass 
 $ \log \kk = \log{(-m_b^2/m_h^2)}$, 
logarithms of the ratio of the transverse momentum  $p_\perp$ divided by the bottom quark mass
$\log{(y\,z/\kk)} =  \log{\left( -p_\perp^2/m_b^2\right)}$ and the logarithms of 
the ratio of two small parameters $u$ and $t$, $\log(y/z) = \log(t/u)$.
To simplify the notation, we define
\begin{equation}
L = \log{(\kk)} = \log \left(\frac{-m_b^2}{ m_h^2}\right),
\;\;\; 
\tau =  \frac{\log{(y\,z/\kk)}}{\log{(\kk)}},
\;\;\;\; \xi = \log \left(\frac{y}{z}\right).
\end{equation}
In the soft limit, $L \gg 1$, while  $\tau$ and $\xi$,  defined above,  are quantities 
of order one.  Expanding the helicity coefficients  in the soft limit and substituting $N_c = 3$, we   find 
\be
\begin{split} 
&{\overline \Omega}^{(1l),\fin}_{+++} =  L^2\left( 1
+ \frac{\tau^2}{2}  \right) + \frac{\pi^2-24}{6}\,,  \\
&{\overline \Omega}_{++-}^{(1l),\fin} =  L^2\left( 1
- \frac{1}{2} \tau^2\right) - \frac{\pi^2+24}{6} \,, 
\end{split} 
\ee
at the one loop and 
\be
\begin{split}
&{\overline \Omega}_{+++}^{(2l),\fin} = 
L^4 \left( \frac{13 \,\tau ^4}{144}+\frac{\tau ^3}{24}-\frac{17 \,\tau ^2}{48}
-\frac{3 \,\tau }{4}-\frac{17}{72}\right)
 \\
&+ L^3 \left(\bar {\beta}_0 
\left(-\frac{\tau ^3}{4}-\frac{\tau ^2}{4}-\frac{\tau }{2}-\frac{1}{2}\right)-\frac{3 \tau
   ^3}{4}+\frac{\tau ^2}{6}-\frac{\tau }{6}+\frac{5}{3}\right) 
 \\
&+ L^2 \left( -\frac{\tau ^2 \, \xi ^2}{48} +\frac{31 \, \pi ^2\,  \tau ^2}{144}+\frac{23\,  \tau ^2}{6}
+\frac{\pi ^2\,  \tau }{72}+3\, \tau
   +\frac{3 \, \xi ^2}{8}-\frac{19\, \pi ^2}{144}
   +\frac{9}{2}\right) 
 \\
&+ L \, \left( \bar {\beta}_0  (\tau+1) \left(-\frac{\pi ^2 }{12}
+2 \right)
-4 \tau  \zeta_3-\frac{\pi
   ^2 \tau }{4}+\frac{2 \tau }{3}+\frac{64 \zeta_3}{3}-\frac{5 \pi ^2}{6}-\frac{52}{3}\right)
 \\
&-\frac{\pi ^2 \xi ^2}{144} -\frac{3 \,\xi ^2}{2}+27 \,\zeta_3 
+\frac{131 \,\pi^4}{270}+\frac{16 \,\pi ^2}{9}-\frac{188}{3}
+ i\, \pi\,\frac{3}{2}\, \bar {\beta}_0\; \Omega_1^{(1l,N_f+1)}\,, 
\end{split} 
\ee
\be
\begin{split} 
&{\overline \Omega}_{++-}^{(2l),\fin} = 
L^4 \left( \frac{3 \,\tau ^4}{16}+\frac{17\, \tau ^3}{72}-\frac{19 \,\tau ^2}{48}
-\frac{3 \,\tau }{4}-\frac{17}{72}\right)
 \\
&+ L^3 \left( \bar {\beta}_0\left(\frac{\tau ^3}{4}+\frac{\tau ^2}{4}
-\frac{\tau }{2}-\frac{1}{2}\right)+\frac{3 \tau
   ^3}{4}-\frac{\tau ^2}{6}-\frac{\tau }{6}+\frac{5}{3}\right) 
 \\
&+ L^2 \left( \frac{\tau ^2 \xi ^2}{48}+\frac{43\, \pi ^2 \tau ^2}{48}
+\frac{19 \, \pi ^2 \tau }{72}+3 \tau +\frac{3 \,\xi ^2}{8}-\frac{7\,
   \pi ^2}{48}+\frac{9}{2}\right) 
 \\
&+ L \, \left( \bar \beta_0 (\tau+1)
\left(\frac{\pi ^2}{12}+2 \right )
-\frac{16 \tau  \zeta_3}{3}+\frac{\pi ^2 \tau }{4}+\frac{2 \tau }{3}+\frac{34 \zeta_3}{3}
-\frac{17 \pi ^2}{18}-\frac{52}{3} \right)
 \\
&+\frac{\pi ^2 \,\xi ^2}{144}-\frac{3\, \xi ^2}{2}+52 \,\zeta_3 +\frac{20 \,\pi ^4}{27}
+\frac{7 \,\pi ^2}{9}-\frac{188}{3}
+ i\, \pi\,  \frac{3}{2} \, \bar {\beta}_0 \Omega_2^{(1l,N_f+1)}\,
\label{eq:softlim2l}
\end{split}
\ee
at the two loops. Note that $\bar {\beta}_0 = 11/2 - 2/3\,T_R\,(N_f+1) $ is the QCD 
$\beta$-function in a theory with $N_f+1$ active flavors.  
We have checked that the abelian 
${\cal O}(L^4)$ part of the soft  limit of the helicity coefficients agrees with the results 
of Ref.\cite{Melnikov:2016emg}; all other terms in that formula are new. 


A second interesting limit is the collinear one. Specifically, 
we are interested in the case when the momenta of gluons $1$ and $3$ become parallel. 
This implies  $m_b^2 \ll t \ll m_h^2 \sim s \sim u $ 
and $y \to 0$, $z \sim {\cal O}(1)$.
Note that in this limit 
 $\log{(y/\kk)} \approx \log{(-t/m_b^2)}$ is considered to be large. 
Similarly to the soft limit, we introduce the notation 
\begin{equation}
\eta = \frac{\log{(y/\kk)}}{\log{(\kk)}}\,.
\end{equation}

At one loop we find 
\be
\begin{split}
&{\overline \Omega}_{+++}^{(1l),\fin} =  L^2\left( \frac{\eta ^2}{2}+1 \right) 
+ L\, \eta\, \left( \log(z)  + \log(1-z)\right)  \\
&\quad + \frac{1}{2} \log ^2(1-z)+\frac{1}{2}\log ^2(z)-\log (z) \log
   (1-z)+\frac{\pi ^2}{6}-4, 
\\
&{\overline \Omega}_{++-}^{(1l),\fin} =  L^2\left( \frac{\eta ^2}{2 (1-z)}-\eta ^2+1 \right)
+ L  \, \eta \, \left( \frac{\log(1-z)-\log(z)}{1-z}\right) \\
&\quad + \frac{2 \text{Li}_2(1-z)-2 \text{Li}_2(z)-8 z-\log ^2(1-z)+\log ^2(z)+8}{2 (z-1)} \,.
\end{split} 
\ee

The complete two-loop result in the $y \to 0$ limit is not sufficiently compact 
to be presented in the paper. For this reason, 
we write the amplitude retaining the coefficients of 
leading and subleading logarithms. We find
\be
\begin{split}
&{\overline \Omega}_{+++}^{(2l),\fin} =  L^4\left(  \frac{5 \eta ^4}{72}+\frac{5}{36} \right)   \\
&\quad+ L^3\, \Bigg[  
\left(\frac{13 \eta ^3}{36}+\frac{\eta ^2}{12}-\frac{3
   \eta }{2}-\frac{3}{2}\right) \left( \log(z) + \log(1-z) \right)  \\
   &\quad + \bar {\beta}_0 \left(-\frac{\eta ^3}{4}-\frac{\eta ^2}{4}
   -\frac{\eta }{2}-\frac{1}{2}\right)-\frac{3 \eta ^3}{4}
   +\frac{\eta ^2}{6}-\frac{\eta }{6}+\frac{5}{3}
   \Bigg] \,,  \\
&\overline{\Omega}_{++-}^{(2l),\fin} =  
L^4\left(  -\frac{5 \eta ^4}{24 (1-z)}+\frac{5 \eta ^4}{12}-\frac{5
   \eta ^3}{18 (1-z)}+\frac{5 \eta ^3}{9}+\frac{5}{36} \right)  \\
&\quad+ L^3  \, \Bigg[
\left(-\frac{3 \eta ^3}{4 (1-z)}+\frac{3 \eta ^3}{2}-\frac{3 \eta ^2}{4
   (1-z)}+\frac{3 \eta ^2}{2}-\frac{3 \eta }{2}-\frac{3}{2}\right)
   \left( \log(z) + \log(1-z) \right)  \\
  &\quad+\frac{5 \eta ^3}{18 (1-z)}\log(1-z)
+\bar {\beta}_0 \left(-\frac{\eta ^3}{4 (1-z)}+\frac{\eta ^3}{2}-\frac{\eta ^2}{4
   (1-z)}+\frac{\eta ^2}{2}-\frac{\eta }{2}-\frac{1}{2}\right)  \\
   &\quad-\frac{3 \eta ^3}{4 (1-z)}+\frac{3 \eta ^3}{2}+\frac{\eta ^2}{6
   (1-z)}-\frac{\eta ^2}{3}-\frac{\eta }{6}+\frac{5}{3}
\Bigg]\,. \label{eq:colllim2l}
\end{split}
\ee
These results show that the structure of large logarithmic corrections 
in the collinear limit is complicated; it does not appear that any one-loop 
structures 
get iterated even at the level of the leading logarithms.

\section{The analytic continuation}
\label{sec:cont}

Up to this point, we studied  the scattering amplitude in the 
decay kinematics. In this case the imaginary part is  generated by an overall
prefactor $(-m_h^2-i\,\delta)^{-n\,\epsilon}$, where $n$ is the number 
of loops.  However, this is not sufficient; indeed, 
to study Higgs boson  production in gluon fusion, we need to  perform an  
analytic continuation of the scattering amplitude. The analytic 
continuation of the $gg \to Hg$ amplitude for the point-like $ggH$ 
interaction vertex was described  in Ref.~\cite{Gehrmann:2002zr}
and we closely follow that paper in our discussion below. 
Following~Ref.~\cite{Gehrmann:2002zr}, we consider three kinematic regions
\begin{align}
&\mbox{region}\;\; (1a)_+\; : \qquad H(p_4) \to g(p_1) + g(p_2)+ g(p_3) 
\, , \label{eq:r1a}
\\
&\mbox{region}\;\; (2a)_+\; : \qquad g(p_1) + g(p_2) \to H(p_4) + g(-p_3)\, ,
\label{eq:r2a}
\\
&\mbox{region}\;\; (4a)_+\; : \qquad g(p_2) + g(p_3) \to H(p_4) + g(-p_1)\,.
\label{eq:r4a}
\end{align}
The region $(1a)_+$ is the decay kinematic region that we considered 
so far in this paper; the region $(2a)_+$ is the kinematic region 
required to describe Higgs production in gluon fusion and 
the region $(4a)_+$ is needed to determine all helicity amplitudes 
for the Higgs production process 
from the two independent ones computed in the previous 
Section. 
In~Ref.~\cite{Gehrmann:2002zr} it was shown how to start from the
helicity amplitudes defined in region $(1a)_+$ and continue them to any other 
kinematic configuration, including regions $(2a)_+$ and $(4a)_+$. 
While the analytic continuation of the spinor structures is 
straightforward, some care has to be taken in continuing the helicity coefficients 
$\Omega_{++ \pm}$, which are written in terms of harmonic and two-dimensional 
harmonic polylogarithms defined in~\cite{Gehrmann:2000zt}.
The analytic continuation of the polylogarithms can be achieved with
a proper change of variables.

In the region $(2a)_+$ the Mandelstam variables 
are 
\be
m_h^2 > 0\,, \quad s>0\,, \quad t,u < 0\,,
\ee
and the analytic continuation from the decay kinematics is performed by 
providing
small and positive imaginary parts to all positive Mandelstam variables.
In particular,  we require $s \to s + i\,\delta$. 
We define two  auxiliary variables
\begin{equation}
u_1 = -\frac{t}{s} = -\frac{y}{1-y-z}\,, \qquad v_1 =  \frac{m_h^2}{s} = \frac{1}{1-y-z}\,,
\label{eq7:5}
\end{equation}
which fulfill the following condition   
\be
0 \leq u_1 \leq v_1\,, \qquad 0 \leq v_1 \leq 1\,.
\ee

In the region $(4a)_+$, instead, we have 
$$m_h^2 > 0\,, \quad s<0\,, \quad t< 0\, \quad u>0,$$
with $u \to u + i\,\delta$. 
Similarly to Eq.(\ref{eq7:5}), we define the following auxiliary variables
\begin{equation}
u_2 = -\frac{t}{u} = -\frac{y}{z}\,, \qquad v_2 =  \frac{m_h^2}{u} = \frac{1}{z}\,,
\end{equation}
which fulfill 
\be
0 \leq u_2 \leq v_2\,, \qquad 0 \leq v_2 \leq 1\,.
\ee

One can show that, by changing the arguments of the polylogarithms from $(y,z)$ to
$(u_1,v_1)$ and $(u_2,v_2)$ in regions $(2a)_+$ and $(4a)_+$ respectively,
and rewriting  the result appropriately, one can extract 
all imaginary
parts explicitly. In this way, the result 
for the helicity amplitudes can be rewritten in terms of explicitly 
real one-
and two-dimensional harmonic polylogarithms 
of the new arguments $u_1,v_1$ and $u_2,v_2$,  respectively. 
The one- and two-loop helicity coefficients 
in the decay kinematics Eq.(\ref{eq:r1a})
and analytically continued to the scattering 
kinematics~Eqs.(\ref{eq:r2a},\ref{eq:r4a})
are available as ancillary files with the arXiv submission of this paper.

\section{Conclusions} \label{sec:conclusions} 

We computed the two-loop helicity amplitudes for the process $H\to ggg$, 
mediated by a quark 
loop, in the approximation when the mass of the quark is small compared to other kinematic  
parameters of the process. The expansion in the small quark mass is used to solve the differential 
equations for the master integrals, while in all other 
steps of the computation no approximations are made. 

The methodological results of this paper 
establish a framework that allows one to expand the   scattering amplitudes 
around  the limit where all or some of the 
particles in the quantum loops  
can be treated as nearly massless.  There are many potential applications of 
this  approach, including  production of the Higgs boson at high 
transverse momentum and the  electroweak corrections in the Sudakov regime. 

On the phenomenological side, 
there are 
several ways in which the results of the computation reported in this paper can be used. 
The two-loop helicity amplitudes can be combined 
with the real emission process $gg \to Hgg$  to study  
modifications of the Higgs boson transverse momentum distribution 
due to the interference of top 
quark and bottom quark loops in the process $pp \to H+j$ in higher orders in QCD.  Such 
a study seems especially worthwhile given the recent proposals to constrain  charm and bottom Yukawa 
couplings from the kinematic distributions of Higgs bosons produced in proton collisions
\cite{Bishara:2016jga,  Soreq:2016rae}.   

Furthermore, it has been recognized  long ago that 
conventional transverse momentum resummation 
framework can not be applied to  the production of the 
Higgs boson through the loop of light quarks  since 
large non-universal double logarithmic corrections are generated by the virtual amplitude 
itself.  Although these double logarithms are of the Sudakov origin, they 
are  
subtle as the process 
requires  the  helicity flip on the {\it soft} quark lines. It would be interesting 
and instructive to understand if the  resummation of 
these Sudakov-like logarithms  can be achieved.  
We expect that the computation of the helicity amplitudes reported in this 
paper and the 
limits of these helicity amplitudes described  in the previous Section, will 
contribute towards 
this goal.

\subsection*{ Acknowledgments}  
We would like to thank A.~Smirnov and 
A.~von Manteuffel for their help with FIRE5 and 
Reduze2, respectively. 
We also acknowledge useful conversations with F.~Caola,
T.~Gehrmann, D.~Kara, H.~Frellesvig.

\newpage 
\bibliographystyle{JHEP}   
\bibliography{Biblio}

\end{document}